\documentclass[letterpaper]{article} 
\usepackage{aaai2026}  
\usepackage{times}  
\usepackage{helvet}  
\usepackage{courier}  
\usepackage[hyphens]{url}  
\usepackage{multirow}
\usepackage{graphicx} 
\usepackage{natbib}  
\usepackage{caption} 
\usepackage{algorithm}
\usepackage{algorithmic}
\usepackage{booktabs}

\usepackage{newfloat}
\usepackage{listings}
\usepackage{amsmath}  
\usepackage{subfigure} 
\usepackage{array} 
\usepackage{placeins}
\usepackage{etoolbox} 

\DeclareCaptionStyle{ruled}{labelfont=normalfont,labelsep=colon,strut=off} 
\floatstyle{ruled}
\newfloat{listing}{tb}{lst}{}
\floatname{listing}{Listing}
\title{RFNNS: Robust Fixed Neural Network Steganography with Universal Text-to-Image Models}

\author{
  Yu Cheng\textsuperscript{\rm 1,2,*},
  Jiuan Zhou\textsuperscript{\rm 1,*},
  Jiawei Chen\textsuperscript{\rm 1},
  Zhaoxia Yin\textsuperscript{\rm 1,\dag},
  Xinpeng Zhang\textsuperscript{\rm 3}
}
\affiliations{
  \textsuperscript{\rm 1}East China Normal University, Shanghai, China\\
  \textsuperscript{\rm 2}Shanghai Innovation Institute, Shanghai, China\\
  \textsuperscript{\rm 3}Fudan University, Shanghai, China\\
  \textsuperscript{*}\,Equal contribution \qquad
  \textsuperscript{\dag}\,Corresponding author
}

\usepackage{bibentry}

\begin{document}

\maketitle
\begin{abstract}



With the rapid development of generative AI, image steganography has garnered widespread attention due to its unique concealment. Recent studies have demonstrated the practical advantages of Fixed Neural Network Steganography (FNNS), notably its ability to achieve stable information embedding and extraction without any additional network training. However, the stego images generated by FNNS still exhibit noticeable distortion and limited robustness. These drawbacks compromise the security of the embedded information and restrict the practical applicability of the method. To address these limitations, we propose Robust Fixed Neural Network Steganography (RFNNS). Specifically, a texture-aware localization technique selectively embeds perturbations carrying secret information into regions of complex textures, effectively preserving visual quality. Additionally, a robust steganographic perturbation generation (RSPG) strategy is designed to enhance the decoding accuracy, even under common and unknown attacks. These robust perturbations are combined with AI-generated cover images to produce stego images. Experimental results demonstrate that RFNNS significantly improves robustness compared to state-of-the-art FNNS methods, achieving an average increase in SSIM of 23\% for recovered secret images under common attacks. Furthermore, the LPIPS value of recovered secrets images against previously unknown attacks achieved by RFNNS was reduced to 39\% of the SOTA method, underscoring its practical value for covert communication. The code is available at https://github.com/edu-yinzhaoxia/RFNNS-Robust-Fixed-Neural-Network-Steganography-with-Universal-Text-to-Image-Models

\end{abstract}


\section{Introduction}

With the rapid development of generative AI, the widespread application of generated content has become increasingly prevalent in daily life, raising significant concerns about data security. 
Steganography \cite{lan2023robust,li2024purified,zhang2021universal,kombrink2024image,meng2025robust}, a critical information hiding technique \cite{yang2023autostegafont,xue2025physical,ji2025speech}, ensures covert communication by embedding secret information in carriers such as images while remaining undetectable to humans and machine eavesdroppers, effectively safeguarding data security. 

Traditional steganography employs simple schemes such as least significant bit (LSB) replacement \cite{van1994digital}. Adaptive steganography selects suitable regions to modify during embedding. Recent advancements in deep neural networks (DNNs) have transformed steganography into a data-driven and learning-based approach \cite{baluja2017hiding,jing2021hinet,chen2022hiding}. However, this method faces two significant challenges: (1) it requires substantial data and computational resources to train effective neural networks; (2) the need to transmit trained models between senders and receivers prior to covert communication not only incurs storage overhead but also heightens the risk of detection by eavesdroppers, thereby compromising security.

\begin{figure}[t]
    \centering
    \includegraphics[width=1.0\linewidth]{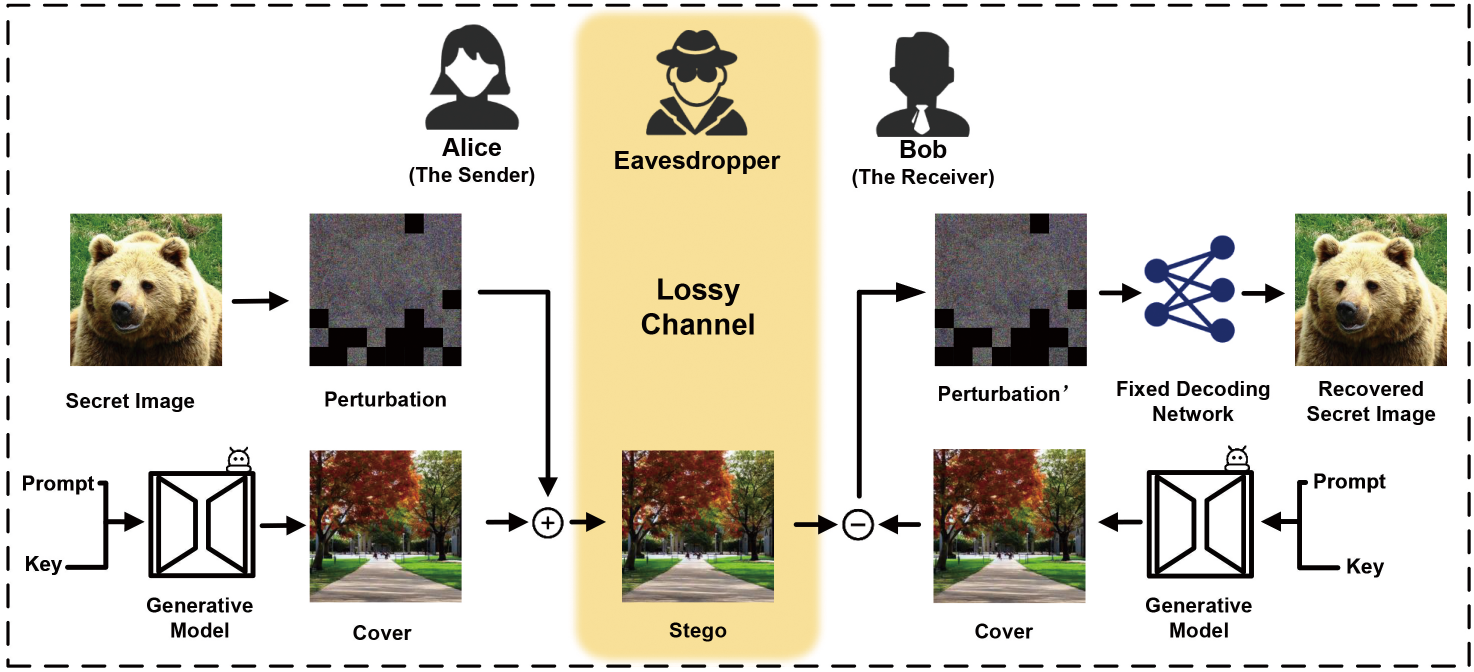}
    \caption{The process of sending and extracting in RFNNS.}
    \label{fig:pipeline}
\end{figure}

To avoid training and transmission of steganographic networks, researchers have employed Fixed Neural Networks (FNNs) \cite{ghamizi2021evasion,kishore2021fixed,luo2023securing,li2024cover} to embed and extract information. This approach leverages adversarial perturbations to modify the cover image such that the stego image can trigger a fixed-parameter decoding network to output the secret information. Covert communication can be achieved by sharing only the fixed decoding network architecture and the random seed to initialize the weights between the sender and the receiver. Nevertheless, existing FNNS methods are currently characterized by poor robustness against common image attacks, low stego image quality, and unsatisfactory anti-steganalysis performance. These limitations severely restrict the further development of this technology.

In response to the aforementioned challenges, we propose an RFNNS method. Unlike previous FNNS methods \cite{ghamizi2021evasion,kishore2021fixed,luo2023securing,li2024cover}, the perturbation embedded in our approach is not global but localized within selected regions. We propose a texture-aware localization technique that introduces perturbations carrying secret information into regions with high textural complexity that are less perceptible to the human eye. In addition, we devise a Robust Steganographic Perturbation Generation (RSPG) strategy that synthesizes perturbations resilient to a variety of common image attacks while keeping the distortion introduced into the stego images imperceptibly low. In practical applications, the receiver employs the shared secret key to access the meticulously designed decoding network we have developed, thereby reliably extracting the secret information. The sending and extraction process is shown in the Fig. \ref{fig:pipeline}.

To evaluate the effectiveness of RFNNS in terms of visual quality, anti-steganalysis performance, and robustness, comprehensive benchmarking experiments were conducted against state-of-the-art FNNS methods. Experimental results indicate that RFNNS consistently achieves better performance compared with all baseline approaches. In particular, RFNNS demonstrates outstanding robustness generalization, maintaining high-quality recovery of secret images even under previously unknown attack scenarios. Experimental results demonstrate that RFNNS significantly improves robustness compared to state-of-the-art FNNS methods, achieving an average increase in SSIM of 23\% for recovered secret images under common attacks. Moreover, under previously unknown attacks, the LPIPS value of recovered secrets achieved by RFNNS was reduced to 39\% of the SOTA method, underscoring its significant robustness advantage. 

Our main contributions are summarized below:

\begin{itemize}
\item A texture‑aware localization technique is proposed to embed perturbations carrying secret information into regions of high texture complexity, which are less perceptible to the human eye. This effectively reduces the distortion of the cover image caused by the perturbations.
\item A RSPG strategy is designed to actively simulate potential attack scenarios that images may encounter during transmission. This strategy ensures that high quality secret images can still be reliably recovered from the stego image even after it has been subjected to common or previously unknown image attacks.
\item Leveraging its meticulously designed fixed decoding network, RFNNS reliably recovers secret images even under common attacks. In addition, it surpasses leading FNNS baselines in visual quality, anti‑steganalysis performance, and robustness.

\end{itemize}

\section{Related Work}
\subsection{Traditional Image Steganography}

Traditional image steganography generally relies on manually designed algorithms to subtly embed secret information into cover images while maintaining their visual quality. Traditional image steganography methods can be broadly classified into spatial domain \cite{chan2004hiding,pan2011image} and transform domain \cite{westfeld2001f5} approaches. To further enhance the undetectability of stego images, researchers have proposed adaptive image steganography techniques \cite{holub2012designing}. Adaptive steganography operates within a distortion coding framework, epitomised by the Syndrome Trellis Codes (STCs) scheme of Filler et al. \cite{filler2011minimizing} and later variants that fine‑tune the distortion metric for different covers \cite{holub2013digital,li2014new}. To remain hidden, these methods cap the payload at roughly 0.5 bpp. Robust steganography aims to resist channel degradations \cite{zeng2023robust,tao2018towards,CHENG2025127598}, yet it still struggles with limited capacity and vulnerability to routine image attacks.

\begin{figure*}[t]
    \centering
    \includegraphics[width=0.85\linewidth]{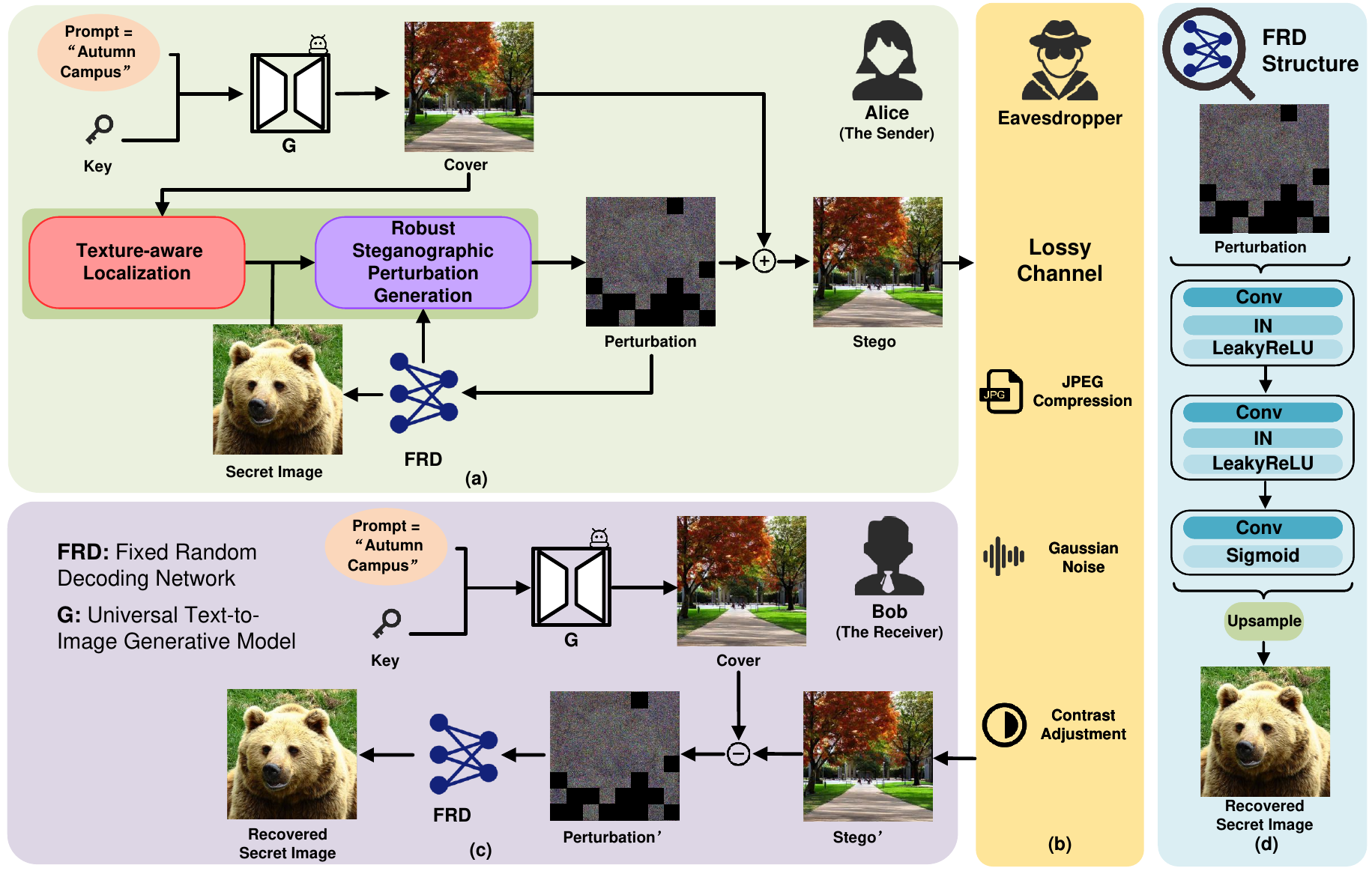}
    \caption{RFNNS framework: (a): Alice (The Sender) employs the proposed texture-aware localization technique to identify embedding regions corresponding to the perturbation. A RSPG strategy is then utilized to incorporate this perturbation into the AI-generated cover image, guided by a shared key, thereby producing the stego image. (b): The eavesdropping and potential image attacks that a stego image may encounter during transmission over a public channel. (c): Bob (The Receiver) first reconstructs the original cover image using the shared key to isolate the perturbation from the stego image. Subsequently, the same decoding network is employed to recover the secret image. (d): Framework of Fixed Random Decoding Network.}
    \label{fig:framework}
\end{figure*}

\subsection{DNN-based Image steganography}

Deep learning image steganography has moved from the pioneering end‑to‑end autoencoder of Zhu et al. \cite{zhu2018hidden}, through the SteganoGAN 6 bpp framework \cite{zhang2019steganogan}, to the recent StegFormer, which embeds multiple secrets in a single cover at up to 96 bpp while preserving high fidelity and robustness \cite{ke2024stegformer}.

However, these methods generally require extensive training data and computational resources, resulting in large network sizes challenging for covert transmission. FNNS emerged to simplify this process, embedding and extracting secret data through adversarial perturbations without additional training. Ghamizi et al. \cite{ghamizi2021evasion} utilized multi-label evasion attacks for secret encoding. Kishore et al. \cite{kishore2021fixed} increased payload by widening the decoder’s output channels and shaping perturbations via an information loss term.
Luo et al. \cite{luo2023securing} added a shared key to align sender and receiver, blocking unauthorized extraction. Li et al. \cite{li2024cover} combined adversarial perturbations with steganographic search optimization. Nonetheless, FNNS techniques commonly face poor robustness against typical image attacks and significant visual distortions, which limits their practicality.

\subsection{Universal Generative Text-to-Image Models}
In recent years, universal text-to-image models—such as Stable Diffusion XL(SDXL) \cite{ICLR2024_081b0806}, Stable Cascade Model \cite{Pernias2024WrstchenAE}, and Latent Diffusion Model \cite{rombach2022high}, have advanced rapidly. Training in large-scale datasets approximates complex data distributions and has been widely used in AIGC, achieving impressive results in computer vision \cite{ho2022cascaded}, natural language processing \cite{brown2020language}, privacy protection \cite{tang2024once}, and biological sciences \cite{zeng2022deep,lai2025deep}. AIGC has also been utilized in information hiding. RFNNS lets the sender and receiver regenerate an identical AI‑generated cover image from a shared key and prompt, then pinpoint high‑texture regions and through an RSPG strategy, embed localized perturbations, yielding a stego image that enhances practical covert communication.

\begin{figure*}[t]
    \centering
    \includegraphics[width=0.8\linewidth]{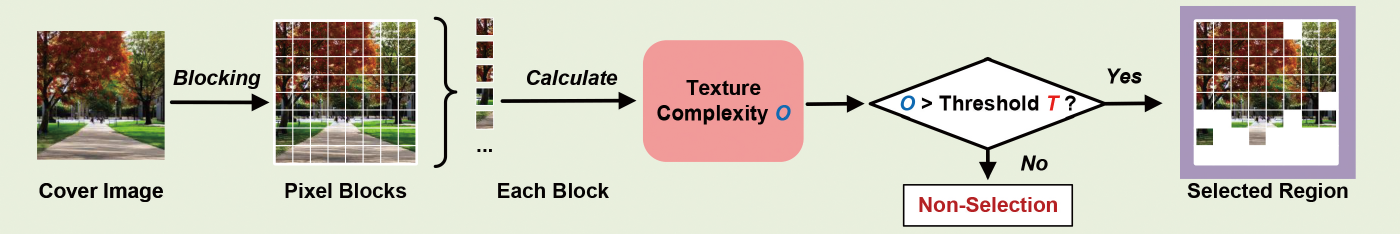}
    \caption{The texture-aware localization technique framework of the proposed method.}
    \label{fig:area}
\end{figure*}

\section{The Proposed Method}
In this section, we first introduce the overall framework of the proposed method. Subsequently, we detail on the texture-aware localization technique and the robust steganographic perturbation generation (RSPG) strategy. Finally, we describe the design of the decoding network.

\subsection{Framework of the Proposed Scheme}

In this study, we propose a novel steganography, called RFNNS. For ease of description, the relevant symbols are shown in Table \ref{tab:notation}. Let $X_c$ represent an AI-generated cover image, with $H_c$ and $W_c$ denoting its height and width, respectively. The secret image to be transmitted, denoted as $S$, is also an RGB image with height $H_s$ and width $W_s$. According to the framework depicted in Fig. \ref{fig:framework}, on the sender side, we input a secret key $k_c$ and a shared \textit{prompt} into a pre-trained universal text-to-image model $G(\cdot)$ to generate the cover image $X_c$.\begin{equation}
\label{eq:ck}
X_c = G(k_{c},prompt)
\end{equation}

\begin{table}[t]
    \centering
    \begin{small}  
    \begin{tabular}{ll}
    \toprule
    \textbf{Notation} & \textbf{Description} \\
    \midrule
    $X_c$            & Cover Image $\in [0,1]^{H_c \times W_c \times 3}$ \\
    $X_s$            & Stego Image $\in [0,1]^{H_c \times W_c \times 3}$ \\
    $\delta$         & Micro Perturbation $\in [0,1]^{H_{\delta} \times W_{\delta} \times 3}$ \\
    $\delta’$         & Recoverd Micro Perturbation $\in [0,1]^{H_{\delta} \times W_{\delta} \times 3}$ \\
    $S$              & Secret Image $\in [0,1]^{H_s \times W_s \times 3}$ \\
    $S’$              & Recoverd Secret Image $\in [0,1]^{H_s \times W_s \times 3}$ \\
    $G(\cdot)$       & Universal Text-to-Image Model \\
    $D_e(\cdot)$     & Decoding Network \\
    \bottomrule
    \end{tabular}
    \caption{Notations}
    \label{tab:notation}
    \end{small}
\end{table}

A texture-aware localization technique is employed to identify embedding regions within the cover image. Subsequently, the secret image is transformed into subtle perturbations denoted as $\delta$ using a RSPG strategy with a fixed decoding network. These perturbations are iteratively updated in response to various potential attacks. The refined robust perturbations are then embedded into predetermined regions of the cover image, ultimately generating the stego image.

On the receiver side, the original cover image is reconstructed using a shared secret key $k_c$ and a shared \textit{prompt}. By comparing this retrieved original image with the received stego image, the receiver extracts perturbation information $\delta'$, which has been subjected to attacks, from the predetermined embedded regions. After sharing the key for the initialization weights $k_w$, the receiver obtains an identical decoding network to that of the sender. By feeding the extracted perturbation $\delta$ into this network, the secret image can be accurately reconstructed from the perturbation $\delta'$. This process can be formally described as: \begin{equation}
\label{eq:des}
\text{De}[k_w](\delta') = S'
\end{equation}

\subsection{Texture-aware Localization}

Existing FNNS methods typically encode secret information by uniformly embedding perturbations throughout the cover image, neglecting the substantial variations in texture complexity among different regions of the image. This uniform embedding strategy often leads to reduced visual quality and degraded overall performance. Embedding perturbations solely in highly textured regions, where human vision is least sensitive, minimizes overall distortion and thus improves visual quality and anti‑steganalysis performance.

As illustrated in Fig. \ref{fig:area}, in practice, the cover image is initially partitioned into multiple equal-sized blocks of dimensions $b_s$ × $b_s$. Subsequently, the texture complexity $O$ is computed for each block, and perturbations are introduced into the blocks whose complexity $O$ exceeds a predefined threshold $T$. We employ the Local Binary Pattern (LBP) \cite{ojala2002multiresolution} method to quantify the $O$ of each block (chosen for its computational simplicity and efficiency, and because it outperformed alternatives in our experiments). For every pixel $p( i , j )$ in an image block, the corresponding LBP value is calculated by comparing the grayscale intensity of the central pixel with its eight neighboring pixels. The binary value $b_k$ for each neighbor pixel $p( i+dy , j+dx )$ is defined as follows: \begin{equation}
\label{eq:bk}
b_k =
\begin{cases}
1, & p(i + dy, j + dx) \ge p(i, j), \\
0, & p(i + dy, j + dx) < p(i, j)
\end{cases}
\end{equation}where $( dy , dx )$ represents the offset of each neighboring pixel relative to the central pixel, and $k$ $(k = 0,1,\dots,7)$ denotes the neighbor index, arranged from left to right and then top to bottom. Following the LBP method described in \cite{pietikainen2010local}, the resulting set of binary values $b_k$ is used to construct an LBP histogram $H(e)$. This histogram is subsequently normalized, yielding the probability distribution $P(e)$, from which we calculate the texture complexity $O$ as the entropy: 

\begin{equation}
O = - \sum_{l=0}^{255} P(e) \log_2 \left[ P(e) + \epsilon \right]
\label{eq:entropy}
\end{equation}where $\varepsilon$ a very small constant is used to avoid undefined values during the logarithmic calculation.

Once the texture complexity $O$ has been calculated for all image blocks, blocks exhibiting $O$ values that exceed the threshold $T$ are marked for perturbation, as shown in the following equation:
 \begin{equation}
\text{perturbation position} =
\begin{cases}
\text{chosen}, & O \geq T \\
\text{unchosen}, & O < T
\end{cases}
\label{eq:perturb_position}
\end{equation}

Using this approach allows us to selectively embed subtle perturbations into blocks with higher texture complexity, thus effectively minimizing the overall perturbation scale.

\subsection{Robust Steganographic Perturbation Generation}

In practical steganography, transmitted images traverse complex and variable channel environments, exposing them to malicious attacks or noise interference that degrade secret information extraction accuracy. To address the aforementioned issues, a RSPG strategy is proposed. We aim to reduce embedding distortion and enhance anti-steganalysis performance through this strategy, while also enabling accurate recovery of the secret image from the stego image after it has undergone various image attacks.

Correspondingly, to mitigate the impact of perturbation on the quality of the cover image, the perturbation introduced during the embedding process should be as minimal as possible. We use a loss function as follows:\begin{equation}
L_1 = \mathrm{MSELoss}(w_p, w_z)
\label{eq:loss1}
\end{equation}
$w_p$ represents the generated perturbation. Here, $w_z$ denotes a zero tensor with the same dimensionality as the perturbation, which guides the perturbation generation process to minimize distortion. Specifically, to constrain the perturbation within the limits, we use $\mu$ to bound $w_p$, as shown in the following formula \ref{eq:constraint}.\begin{equation}
\ w_p \leq \mu
\label{eq:constraint}
\end{equation}

In addition to maintaining image quality, robust extraction of secret information is critical. To accurately recover the embedded data, a second loss function is introduced:\begin{equation}
L_2 = \text{MSELoss}(\text{$S'$}, \text{$S$})
\label{eq:loss2}
\end{equation}

Furthermore, by simulating various attacks during the adversarial noise generation process, a loss function is designed:\begin{equation}
L_3 = \text{MSELoss}(\text{$attack\_S'$}, \text{$S$})
\label{eq:loss3}
\end{equation}

  \begin{equation}
\text{$attack\_S'$} =
\begin{cases}
\text{JPEG\_Compression}(S,QF) \\
\text{Gaussian\_Noise}(S,\rho) \\
\text{Contrast\_Adjustment}(S,\eta) \\
\text{Other Attack}(S,\phi)
\end{cases}
\label{eq:attack_variants}
\end{equation}Where $QF$, $\rho$, $\eta$, and $\phi$ denote the hyperparameters for the respective attack types. This loss function actively simulates potential attacks during the perturbation generation process, thereby effectively enhancing the perturbation’s robustness against common image attacks.

During the perturbation optimization process, we incorporate pre-trained steganalyzers into the later iterations to provide gradient feedback for perturbation refinement, thereby enhancing the anti-steganalysis performance of the generated stego images.  Consequently, the following loss function is formulated:\begin{equation}
L_4 = \text{CE Loss}(\text{$X_s$}, \text{$Label$})
\label{eq:loss4}
\end{equation}

\begin{equation}
\label{eq:crossentropy_single}
\text{CE Loss}(X_{s, y})
= -\log\Biggl(\frac{\exp(X_{s, y})}
                   {\exp(X_{s, 0}) + \exp(X_{s, 1})}\Biggr)
\end{equation}“CE Loss” stands for “CrossEntropyLoss.” $Label$ denotes the classification result provided by the steganalyzer. $y$ denotes the current index, taking values in \{0, 1\}. $X_{s,0}$ represents the logit corresponding to the classification of the image as stego image, and $X_{s,1}$ represents the logit corresponding to the classification of the image as normal.

In practice, we prioritized the visual quality of the stego images by adjusting the weight $L$\textsubscript{1}. Empirical observations suggest that when $L$\textsubscript{1} is reduced to a threshold $Y$, the image distortion introduced to stego images can be almost ignored, thus preserving high visual fidelity. The refined loss function accordingly takes the following form:\begin{equation}
\text{\textit{L}} =
\begin{cases}
 Y + \beta \cdot \text{$L$}_2 + (1 - \beta) \cdot \text{$L$}_3+ \gamma\cdot\text{$L$}_4, & \text{if } \text{$L$}_1 < Y \\
\alpha \cdot \text{$L$}_1 + \beta \cdot \text{$L$}_2 + (1 - \beta) \cdot \text{$L$}_3+ \gamma\cdot\text{$L$}_4, & \text{if } \text{$L$}_1 \geq Y
\end{cases}
\label{eq:adaptive_loss}
\end{equation}
where $\alpha$, $\beta$, and $\gamma$ are hyperparameters that balance the contributions of different loss functions.


The proposed strategy iteratively refines perturbations, preserving the visual quality of both stego and recovered secret images under common and previously unknown attacks. Experimental results demonstrate that the RSPG strategy exhibits remarkable robustness and meets the requirements for covert communication in practical scenarios.


\begin{figure*}[t]
  \centering
  \includegraphics[width=.255\linewidth]{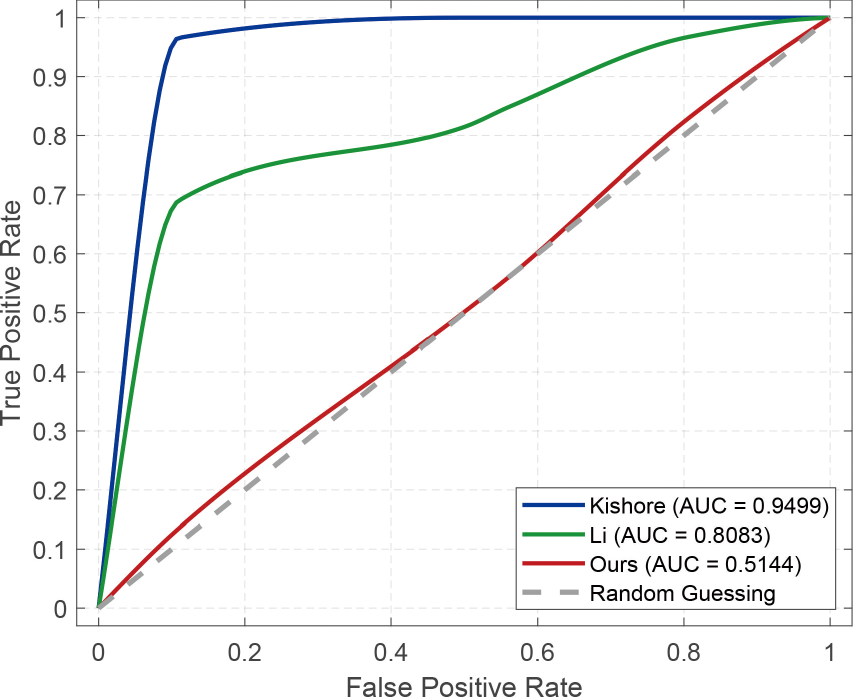}\hfill
  \includegraphics[width=.255\linewidth]{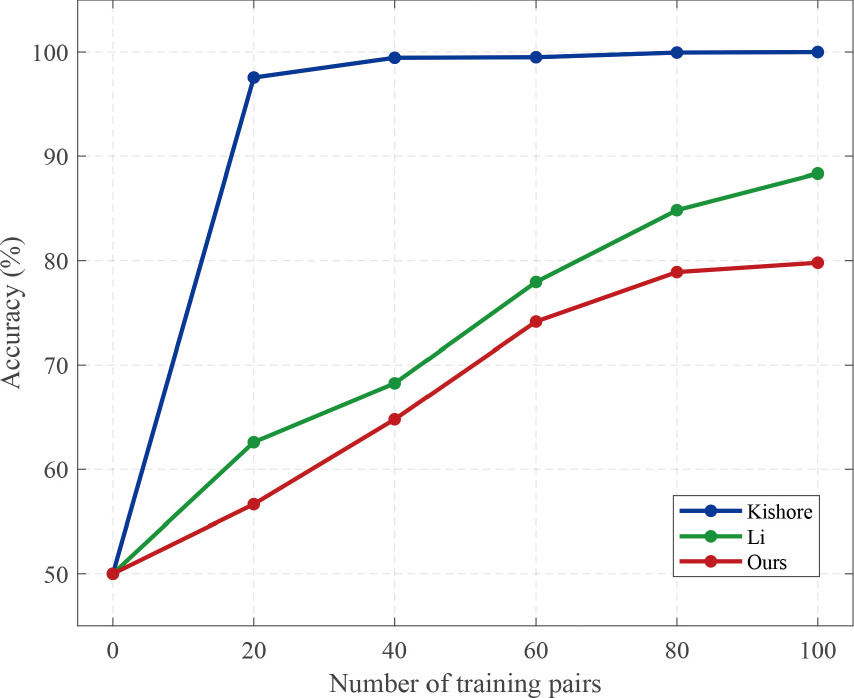}\hfill
  \includegraphics[width=.255\linewidth]{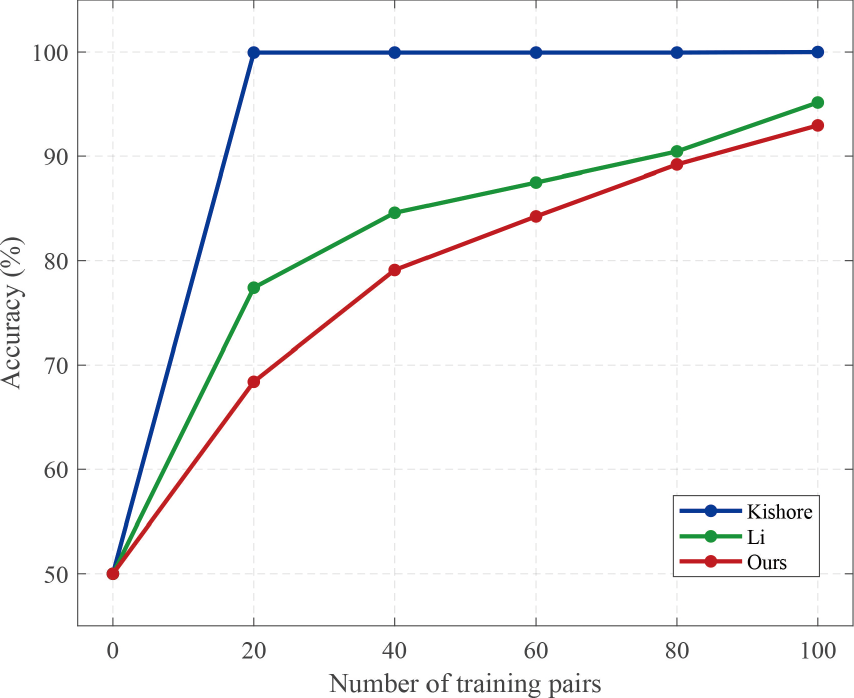}
  \caption{Anti-steganalysis performance at the low embedding capacity:
  (a) StegExpose; (b) YeNet; (c) SiaStegNet.}
  \label{fig:fenxi_low}
\end{figure*}

\begin{table*}[htbp]
  \centering
  \begin{small}          
  \begin{tabular}{@{}l l l l l l l l l l l l l@{}}   
    \toprule[0.8pt]
    \multirow{2}{*}{Capacity} & \multirow{2}{*}{Attack} & \multirow{2}{*}{Factor} &
      \multicolumn{3}{c}{Kishore et al.} &
      \multicolumn{3}{c}{Li et al.} &
      \multicolumn{3}{c}{Ours} \\  
    \cmidrule(lr){4-6}\cmidrule(lr){7-9}\cmidrule(lr){10-12}
    & & & PSNR$\uparrow$ & SSIM$\uparrow$ & LPIPS$\downarrow$
        & PSNR$\uparrow$ & SSIM$\uparrow$ & LPIPS$\downarrow$
        & PSNR$\uparrow$ & SSIM$\uparrow$ & LPIPS$\downarrow$ \\
    \midrule
    \multirow{6}{*}{1.5 bpp} &
      \multicolumn{2}{l}{No Attack} & 24.22 & 0.675 & 0.223 &
        41.17 & \textbf{0.981} & \textbf{0.003} &
        \textbf{41.48} & 0.980 & \textbf{0.003} \\
    \cmidrule{2-12}
    & JPEG Compression & \textit{QF}=80 &
        13.96 & 0.210 & 1.061 &
        23.00 & 0.568 & 0.350 &
        \textbf{25.43} & \textbf{0.703} & \textbf{0.147} \\
    \cmidrule{2-12}
    & Gaussian Noise & $\rho$=0.07 &
        13.93 & 0.193 & 0.890 &
        20.72 & 0.471 & 0.323 &
        \textbf{26.72} & \textbf{0.748} & \textbf{0.124} \\
    \cmidrule{2-12}
    & Contrast Adj. & $\eta$=0.7 &
        12.97 & 0.405 & 0.617 &
        24.87 & 0.885 & \textbf{0.034} &
        \textbf{32.60} & \textbf{0.889} & 0.047 \\
    \midrule
    \multirow{6}{*}{6 bpp} &
      \multicolumn{2}{l}{No Attack} &
        18.98 & 0.577 & 0.393 &
        41.79 & 0.981 & 0.004 &
        \textbf{42.95} & \textbf{0.984} & \textbf{0.003} \\
    \cmidrule{2-12}
    & JPEG Compression & \textit{QF}=80 &
        11.52 & 0.195 & 1.115 &
        21.52 & 0.507 & 0.371 &
        \textbf{21.58} & \textbf{0.565} & \textbf{0.222} \\
    \cmidrule{2-12}
    & Gaussian Noise & $\rho$=0.07 &
        19.13 & 0.584 & 0.392 &
        19.88 & 0.438 & 0.325 &
        \textbf{26.19} & \textbf{0.738} & \textbf{0.130} \\
    \cmidrule{2-12}
    & Contrast Adj. & $\eta$=0.7 &
        13.10 & 0.421 & 0.596 &
        22.85 & 0.758 & \textbf{0.082} &
        \textbf{28.15} & \textbf{0.784} & 0.093 \\
    \bottomrule[1.2pt]
  \end{tabular}
  \end{small}
  \caption{Stego image quality under different embedding capacities and attack conditions ($\uparrow$ higher is better, $\downarrow$ lower is better).}
  \label{table2}
\end{table*}

\subsection{Decoding Network Construction}

The architecture of the decoding network significantly impacts decoding performance. Prior research \cite{kishore2021fixed,luo2023securing,li2024cover} has shown that architectural choices directly affect the efficacy of decoding. As illustrated in Fig. \ref{fig:framework}(d), our proposed network integrates convolutional (Conv) layers, instance normalization (IN), LeakyReLU activations, and a final sigmoid activation. Each Conv layer contains parameters structured as four-dimensional tensors, with fixed kernel sizes of 3. To finely adjust embedding capacity, Conv layers with varied strides are strategically used. Adjusting these strides directly alters the spatial relationship between secret information ($\delta$/$S$) and the cover image ($X_c$), allowing precise control over embedding capacities. Once the decoding network $D[\cdot]$ is established using the shared key $k_w$, both the sender and the receiver independently replicate identical networks, greatly reducing the necessary information exchange. This enhances the security and practicality of the steganographic algorithm.

\section{Experiments}

This section presents the experimental setup and results. Section \ref{settings} describes the setup; Sections \ref{security} and \ref{robustness} report security and robustness, respectively. Appendix A provides ablations. Appendices B and D extend robustness to additional attacks and analyze performance across capacities. Appendix E and F covers text-to-image model selection and texture-complexity experiments, and Appendix G summarizes computational efficiency and hyperparameter choices.

\subsection{Experimental Settings}\label{settings}

{\bfseries Datasets.} We employ a pre-trained Stable Diffusion model \cite{rombach2022high} as the text-to-image model \(G(\cdot)\) to construct a cover image dataset comprising 3,000 images, each with a resolution of \(512 \times 512\) pixels. Each image is generated using a unique seed \(k_c\) and a fixed textual prompt "Campus". The dataset is evenly split into three 1,000-image subsets, each used to embed secret images randomly drawn from COCO \cite{lin2014microsoft}, CelebA \cite{liu2015deep}, and ImageNet \cite{russakovsky2015imagenet}, respectively. The secret images are resized to \(256 \times 256\) and \(128 \times 128\) pixels to accommodate high (6 bpp) and low (1.5 bpp) embedding capacities. For an embedding capacity of 1.5 bpp, the decoding network employs a convolutional kernel size of 84; for 6 bpp, the kernel size is increased to 104. 

{\bfseries Hyperparameters.} 
Experiments showed that our method performs best when texture complexity is evaluated on $8 \times 8$ blocks; hence, we fix the block size at $b_s = 8$ in all subsequent experiments. To facilitate optimization, the dimensionality of the perturbation is ensured to be no smaller than that of the secret image, allowing for more effective information extraction. Following the approach of Cs-FNNS, the total number of optimization iterations is set to 1,500. The initial learning rate is $1 \times 10^{-1.25}$, and it is halved every 500 iterations. The perturbation bound $\mu$ is fixed at 0.2. After 1,400 iterations, we incorporate pre-trained steganalysis networks, including SRNet \cite{boroumand2018deep} and SiaStegNet \cite{you2020siamese}, to provide gradient feedback for further perturbation refinement. According to our experiments, when $\text{$L$}_1$ in Equation \ref{eq:adaptive_loss} drops below 0.001, the perturbations generated have negligible impact on the visual quality of the stego image. Therefore, we set the threshold $Y = 0.001$. In attack-free scenarios, the parameters in Equation \ref{eq:adaptive_loss} are configured as $\beta = 3$ and $\text{$L$}_3 = 0$, focusing optimization on information recovery. In contrast, under attack conditions, $\beta$ is dynamically reduced to 0.5 to balance robustness and recovery. The remaining hyperparameters $\alpha$ and $\gamma$ are empirically fixed at 1 and $1 \times 10^{-5}$, respectively, to ensure stable convergence while preserving secret image integrity. In Equation \ref{eq:perturb_position}, the threshold $T$ for texture complexity is set to 4.5. We recommend that the receiver use a lightweight post-processing denoising technique described in \cite{zhang2017beyond} to enhance the quality of recovered secret images.

\begin{figure}[t]
  \centering
  \includegraphics[width=\linewidth]{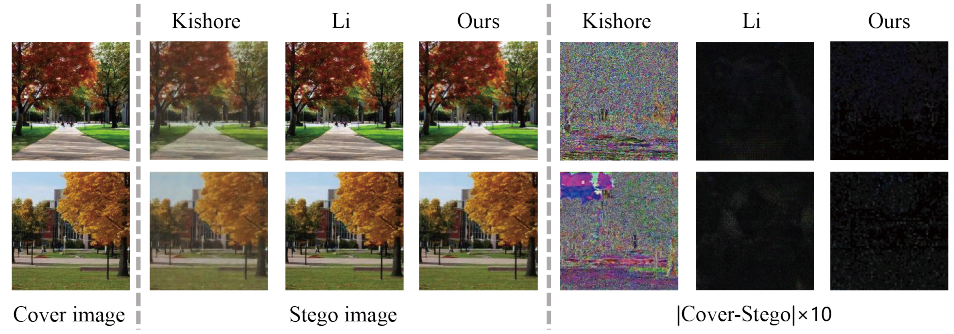}
  \caption{The image quality of stegos for different methods.}
  \label{fig:cancha}
\end{figure}

\subsection{Security}
\label{security}
In image steganography, security is typically categorized into imperceptibility and anti-steganalysis performance.

\noindent 4.2.1 \quad \textit{Imperceptibility}. Image quality is a critical metric for evaluating the imperceptibility. Fig. \ref{fig:cancha} provides a comparative visualization between the RFNNS and two other methods in terms of the quality of recovered secret images. It is evident that the stego images generated by RFNNS are nearly indistinguishable from their respective cover images, as indicated by the almost invisible residuals magnified by a factor of 10. This result demonstrates that the proposed method preserves high chromatic fidelity while introducing only negligible perceptible artifacts. As shown in Table \ref{table2}, the stego images generated by RFNNS surpass those produced by other FNNS methods under both attacked and attack‑free conditions. In particular, the proposed method achieves superior PSNR values in nearly all test cases. Under a 6 bpp embedding rate and a Gaussian noise condition with a variance of 0.07, the SSIM improvement reaches 68. 5\%, while the LPIPS metric is reduced to as low as 40\% of the score achieved by the best competing method, highlighting the improved perceptual fidelity of the stego images.

\begin{table*}[htbp]
  \centering
  \begin{small}          
  \begin{tabular}{@{}l l l l l l l l l l l l l@{}} 
    \toprule[0.8pt]
    \multirow{2}{*}{Capacity} & \multirow{2}{*}{Attack} & \multirow{2}{*}{Factor} &
      \multicolumn{3}{c}{Kishore et al.} &
      \multicolumn{3}{c}{Li et al.} &
      \multicolumn{3}{c}{Ours} \\  
    \cmidrule(lr){4-6}\cmidrule(lr){7-9}\cmidrule(lr){10-12}
    & & & PSNR$\uparrow$ & SSIM$\uparrow$ & LPIPS$\downarrow$
        & PSNR$\uparrow$ & SSIM$\uparrow$ & LPIPS$\downarrow$
        & PSNR$\uparrow$ & SSIM$\uparrow$ & LPIPS$\downarrow$ \\
    \midrule

    \multirow{6}{*}{1.5 bpp} &
      \multicolumn{2}{l}{No Attack} & 33.43 & 0.922 & 0.056 &
        \textbf{35.34} & \textbf{0.949} & 0.019 &
        34.14 & 0.943 & \textbf{0.017} \\
    \cmidrule{2-12}
    & JPEG Compression & \textit{QF}=80 &
        12.14 & 0.263 & 0.642 &
        27.38 & 0.840 & 0.073 &
        \textbf{29.27} & \textbf{0.858} & \textbf{0.072} \\
    \cmidrule{2-12}
    & Gaussian Noise & $\rho$ = 0.07 &
        14.53 & 0.435 & 0.479 &
        23.62 & 0.753 & \textbf{0.145} &
        \textbf{26.08} & \textbf{0.756} & 0.169 \\
    \cmidrule{2-12}
    & Contrast Adj. & $\eta$ = 0.7 &
        12.06 & 0.235 & 0.618 &
        13.86 & 0.363 & 0.611 &
        \textbf{33.68} & \textbf{0.950} & \textbf{0.019} \\
    \midrule

    \multirow{6}{*}{6 bpp} &
      \multicolumn{2}{l}{No Attack} &
        15.69 & 0.472 & 0.491 &
        \textbf{34.61} & \textbf{0.938} & \textbf{0.027} &
        31.09 & 0.910 & 0.058 \\
    \cmidrule{2-12}
    & JPEG Compression & \textit{QF}=80 &
        11.55 & 0.223 & 0.695 &
        18.45 & 0.651 & 0.311 &
        \textbf{22.85} & \textbf{0.696} & \textbf{0.260} \\
    \cmidrule{2-12}
    & Gaussian Noise & $\rho$ = 0.07 &
        14.23 & 0.406 & 0.542 &
        19.07 & 0.643 & 0.296 &
        \textbf{24.49} & \textbf{0.665} & \textbf{0.294} \\
    \cmidrule{2-12}
    & Contrast Adj. & $\eta$ = 0.7 &
        13.27 & 0.272 & 0.624 &
        14.19 & 0.406 & 0.652 &
        \textbf{28.69} & \textbf{0.879} & \textbf{0.071} \\
    \bottomrule[1.2pt]
  \end{tabular}
  \end{small}
  \caption{Recovered secret image quality under different embedding capacities and attack conditions.}
  \label{tab3}
\end{table*}

\noindent 4.2.2 \quad \textit{Anti-steganalysis Performance}. Following the protocol of Luo et al. \cite{luo2023securing}, we fed 3000 cover-stego pairs to StegExpose and plotted the ROC curves in Fig. \ref{fig:fenxi_low}(a). The RFNNS curve coincides with the diagonal 'random-guess' line, whereas competing methods deviate markedly, indicating that RFNNS offers the lowest detectability.

For CNN‑based detectors YeNet,\cite{ye2017deep} and SiaStegNet \cite{you2020siamese}, the same 3000 pairs were split into 2000 for training and 1000 for testing, and the training subset was gradually expanded following the scheme of Guan et al. \cite{guan2022deepmih} and Jing et al. \cite{jing2021hinet}. Across all training sizes (Fig.\ref{fig:fenxi_low} (b), (c)), RFNNS remains the hardest target: at 1.5 bpp with only 100 training pairs, YeNet reaches no more than 80\% detection accuracy and SiaStegNet 92. 95\%, both noticeably lower than for the baselines. Due to space limitations, results on anti-steganalysis performance at a high embedding capacity (6 bpp) are presented separately in Appendix Section C. These results confirm that RFNNS preserves its advantage across payloads and training regimes.

The RFNNS outperforms existing FNNS methods in terms of imperceptibility and anti-steganalysis performance. 
This advantage comes from the texture‑aware localization technique, which confines perturbation‑induced distortions to minimal regions. Moreover, the RSPG strategy further ensures that the discrepancy between the stego image and its cover is kept to a low level.

\subsection{Robustness}
\label{robustness}
\noindent 4.3.1 \quad \textit{Robustness under non-attack conditions}. Table \ref{tab3} presents the visual quality metrics for recovered secret images generated by different methods. Under non-attack conditions, the performance of RFNNS is largely consistent with SOTA methods at 1.5 bpp. In the higher capacity scenario, RFNNS maintains an SSIM value greater than 0.9, demonstrating that it continues to achieve satisfactory quality in terms of hidden information extraction.

\noindent 4.3.2 \quad \textit{Robustness with attack conditions}. The stego image transmitted over communication channels inevitably faces diverse and unpredictable interference. These attacks can compromise the accuracy of secret information extraction, thereby undermining the practical reliability of covert communication systems. This section provides a comprehensive robustness evaluation of existing FNNS methods against common image attacks, taking three attacks as representative examples. As shown in Table \ref{tab3}, the proposed RFNNS method consistently outperforms existing FNNS approaches under both low and high embedding capacities across various attack scenarios. For instance, under contrast adjustment attacks, RFNNS achieves approximately 15 dB higher PSNR, nearly doubles the SSIM, and reduces the LPIPS value to 10\% compared to state-of-the-art methods.

As further demonstrated in Table \ref{tab4}, the perturbations optimized by the RSPG strategy exhibit strong generalization capabilities, effectively handling previously unknown attacks. Specifically, RFNNS improves the PSNR of recovered secret images by around 34\%. Additional robustness evaluations of RFNNS under other common image attacks and its generalization performance against unknown attacks are provided in the appendix Section B. These notable improvements primarily result from the RSPG strategy, which enhances the generalization capability of perturbation robustness by simulating various attack scenarios during optimization. In contrast, the method proposed by Li et al. \cite{li2024cover} applies global perturbations uniformly to the entire cover image, inherently limiting robustness enhancement. Additionally,the approach of Li et al. incorporates simulated attacks only once every two optimization iterations, leading to unstable optimization loss and, consequently, hindering convergence toward robust perturbations.

\begin{table}[t]
  \centering
  \begin{small} 
  \begin{tabular}{@{}l@{\hspace{3pt}}l@{\hspace{3pt}}
      *{6}{>{\centering\arraybackslash}c@{\hspace{2pt}}}@{}}
    \toprule
    \multirow{2}{*}{Type} & \multirow{2}{*}{Capacity} &
      \multicolumn{3}{c}{Li et al.} &
      \multicolumn{3}{c}{Ours} \\
    \cmidrule(r){3-5}\cmidrule(l){6-8}
      & & PSNR$\uparrow$ & SSIM$\uparrow$ & LPIPS$\downarrow$
        & PSNR$\uparrow$ & SSIM$\uparrow$ & LPIPS$\downarrow$ \\
    \midrule
    Type1 & 1.5bpp & 28.64 & 0.865 & 0.063 & \textbf{31.47} & \textbf{0.931} & \textbf{0.029}\\
    Type2 & 1.5bpp & 24.43 & 0.806 & 0.104 & \textbf{31.52} & \textbf{0.935} & \textbf{0.027}\\
    \midrule
    Type1 & 6bpp   & 19.93 & 0.695 & 0.241 & \textbf{28.29} & \textbf{0.853} & \textbf{0.077}\\
    Type2 & 6bpp   & 16.25 & 0.559 & 0.401 & \textbf{28.00} & \textbf{0.847} & \textbf{0.084}\\
    \bottomrule
  \end{tabular}
  \end{small}
  \caption{Recovered secret image quality under different embedding capacities
         and unknown attack conditions. Type 1 simulates JPEG compression
         and contrast adjustment, with Gaussian noise as the actual attack;
         Type 2 simulates JPEG compression, image scaling, and contrast
         adjustment, with Gaussian noise as the actual attack.}
  \label{tab4}
\end{table}

\section{Conclusion}

In this paper, we propose a RFNNS that combines robust perturbations carrying secret information with AI-generated cover images to produce stego images. The introduced texture-aware localization technique effectively enhances the security of steganography. Additionally, a designed RSPG strategy provides significant robustness against various common image attacks. Experimental results confirm that the proposed method surpasses existing approaches at both low and high embedding capacities, while still maintaining high‐fidelity recovery of secret images even against unknown attacks.

\bibliography{aaai2026}

@article{filler2011minimizing,
  title={Minimizing additive distortion in steganography using syndrome-trellis codes},
  author={Filler, Tom{\'a}{\v{s}} and Judas, Jan and Fridrich, Jessica},
  journal={IEEE Transactions on Information Forensics and Security},
  volume={6},
  number={3},
  pages={920--935},
  year={2011},
  publisher={IEEE}
}

@inproceedings{holub2012designing,
  title={Designing steganographic distortion using directional filters},
  author={Holub, Vojt{\v{e}}ch and Fridrich, Jessica},
  booktitle={2012 IEEE International workshop on information forensics and security (WIFS)},
  pages={234--239},
  year={2012},
  organization={IEEE}
}

@inproceedings{holub2013digital,
  title={Digital image steganography using universal distortion},
  author={Holub, Vojt{\v{e}}ch and Fridrich, Jessica},
  booktitle={Proceedings of the first ACM workshop on Information hiding and multimedia security},
  pages={59--68},
  year={2013}
}

@inproceedings{li2014new,
  title={A new cost function for spatial image steganography},
  author={Li, Bin and Wang, Ming and Huang, Jiwu and Li, Xiaolong},
  booktitle={2014 IEEE International conference on image processing (ICIP)},
  pages={4206--4210},
  year={2014},
  organization={IEEE}
}

@inproceedings{jing2021hinet,
  title={Hinet: Deep image hiding by invertible network},
  author={Jing, Junpeng and Deng, Xin and Xu, Mai and Wang, Jianyi and Guan, Zhenyu},
  booktitle={Proceedings of the IEEE/CVF international conference on computer vision},
  pages={4733--4742},
  year={2021}
}

@article{chen2022hiding,
  title={Hiding images in deep probabilistic models},
  author={Chen, Haoyu and Song, Linqi and Qian, Zhenxing and Zhang, Xinpeng and Ma, Kede},
  journal={Advances in Neural Information Processing Systems},
  volume={35},
  pages={36776--36788},
  year={2022}
}

@inproceedings{li2024purified,
  title={Purified and unified steganographic network},
  author={Li, Guobiao and Li, Sheng and Luo, Zicong and Qian, Zhenxing and Zhang, Xinpeng},
  booktitle={Proceedings of the IEEE/CVF conference on computer vision and pattern recognition},
  pages={27569--27578},
  year={2024}
}

@inproceedings{ghamizi2021evasion,
  title={Evasion attack steganography: Turning vulnerability of machine learning to adversarial attacks into a real-world application},
  author={Ghamizi, Salah and Cordy, Maxime and Papadakis, Mike and Le Traon, Yves},
  booktitle={Proceedings of the IEEE/CVF International conference on computer vision},
  pages={31--40},
  year={2021}
}

@article{zhang2017beyond,
  title={Beyond a gaussian denoiser: Residual learning of deep cnn for image denoising},
  author={Zhang, Kai and Zuo, Wangmeng and Chen, Yunjin and Meng, Deyu and Zhang, Lei},
  journal={IEEE transactions on image processing},
  volume={26},
  number={7},
  pages={3142--3155},
  year={2017},
  publisher={IEEE}
}

@inproceedings{kishore2021fixed,
  title={Fixed neural network steganography: Train the images, not the network},
  author={Kishore, Varsha and Chen, Xiangyu and Wang, Yan and Li, Boyi and Weinberger, Kilian Q},
  booktitle={International conference on learning representations},
  year={2021}
}

@inproceedings{luo2023securing,
  title={Securing fixed neural network steganography},
  author={Luo, Zicong and Li, Sheng and Li, Guobiao and Qian, Zhenxing and Zhang, Xinpeng},
  booktitle={Proceedings of the 31st ACM international conference on multimedia},
  pages={7943--7951},
  year={2023}
}

@inproceedings{li2024cover,
  title={Cover-separable Fixed Neural Network Steganography via Deep Generative Models},
  author={Li, Guobiao and Li, Sheng and Qian, Zhenxing and Zhang, Xinpeng},
  booktitle={Proceedings of the 32nd ACM International Conference on Multimedia},
  pages={10238--10247},
  year={2024}
}

@inproceedings{lan2023robust,
  title={Robust image steganography: hiding messages in frequency coefficients},
  author={Lan, Yuhang and Shang, Fei and Yang, Jianhua and Kang, Xiangui and Li, Enping},
  booktitle={Proceedings of the AAAI conference on artificial intelligence},
  volume={37},
  number={12},
  pages={14955--14963},
  year={2023}
}

@inproceedings{zhang2021universal,
  title={Universal adversarial perturbations through the lens of deep steganography: Towards a fourier perspective},
  author={Zhang, Chaoning and Benz, Philipp and Karjauv, Adil and Kweon, In So},
  booktitle={Proceedings of the AAAI conference on artificial intelligence},
  volume={35},
  number={4},
  pages={3296--3304},
  year={2021}
}

@article{kombrink2024image,
  title={Image steganography approaches and their detection strategies: A survey},
  author={Kombrink, Meike Helena and Geradts, Zeno Jean Marius Hubert and Worring, Marcel},
  journal={ACM Computing Surveys},
  volume={57},
  number={2},
  pages={1--40},
  year={2024},
  publisher={ACM New York, NY}
}

@article{chan2004hiding,
  title={Hiding data in images by simple LSB substitution},
  author={Chan, Chi-Kwong and Cheng, Lee-Ming},
  journal={Pattern recognition},
  volume={37},
  number={3},
  pages={469--474},
  year={2004},
  publisher={Elsevier}
}

@inproceedings{pan2011image,
  title={Image steganography method based on PVD and modulus function},
  author={Pan, Feng and Li, Jun and Yang, Xiaoyuan},
  booktitle={2011 International Conference on Electronics, Communications and Control (ICECC)},
  pages={282--284},
  year={2011},
  organization={IEEE}
}

@inproceedings{westfeld2001f5,
  title={F5—a steganographic algorithm: High capacity despite better steganalysis},
  author={Westfeld, Andreas},
  booktitle={International workshop on information hiding},
  pages={289--302},
  year={2001},
  organization={Springer}
}

@inproceedings{zhu2018hidden,
  title={Hidden: Hiding data with deep networks},
  author={Zhu, Jiren and Kaplan, Russell and Johnson, Justin and Fei-Fei, Li},
  booktitle={Proceedings of the European conference on computer vision (ECCV)},
  pages={657--672},
  year={2018}
}

@article{zhang2019steganogan,
  title={SteganoGAN: High capacity image steganography with GANs},
  author={Zhang, Kevin Alex and Cuesta-Infante, Alfredo and Xu, Lei and Veeramachaneni, Kalyan},
  journal={arXiv preprint arXiv:1901.03892},
  year={2019}
}

@article{baluja2017hiding,
  title={Hiding images in plain sight: Deep steganography},
  author={Baluja, Shumeet},
  journal={Advances in neural information processing systems},
  volume={30},
  year={2017}
}

@inproceedings{rombach2022high,
  title={High-resolution image synthesis with latent diffusion models},
  author={Rombach, Robin and Blattmann, Andreas and Lorenz, Dominik and Esser, Patrick and Ommer, Bj{\"o}rn},
  booktitle={Proceedings of the IEEE/CVF conference on computer vision and pattern recognition},
  pages={10684--10695},
  year={2022}
}

@inproceedings{Pernias2024WrstchenAE,
  title={W{\"u}rstchen: An Efficient Architecture for Large-Scale Text-to-Image Diffusion Models},
  author={Pablo Pernias and Dominic Rampas and Mats L. Richter and Christopher Joseph Pal and Marc Aubreville},
  booktitle={International Conference on Learning Representations},
  year={2024},
  url={https://api.semanticscholar.org/CorpusID:271532752}
}

@inproceedings{ke2024stegformer,
  title={Stegformer: Rebuilding the glory of autoencoder-based steganography},
  author={Ke, Xiao and Wu, Huanqi and Guo, Wenzhong},
  booktitle={Proceedings of the AAAI Conference on Artificial Intelligence},
  volume={38},
  number={3},
  pages={2723--2731},
  year={2024}
}

@inproceedings{van1994digital,
  title={A digital watermark},
  author={Van Schyndel, Ron G and Tirkel, Andrew Z and Osborne, Charles F},
  booktitle={Proceedings of 1st international conference on image processing},
  volume={2},
  pages={86--90},
  year={1994},
  organization={IEEE}
}

@article{ho2022cascaded,
  title={Cascaded diffusion models for high fidelity image generation},
  author={Ho, Jonathan and Saharia, Chitwan and Chan, William and Fleet, David J and Norouzi, Mohammad and Salimans, Tim},
  journal={Journal of Machine Learning Research},
  volume={23},
  number={47},
  pages={1--33},
  year={2022}
}

@article{brown2020language,
  title={Language models are few-shot learners},
  author={Brown, Tom and Mann, Benjamin and Ryder, Nick and Subbiah, Melanie and Kaplan, Jared D and Dhariwal, Prafulla and Neelakantan, Arvind and Shyam, Pranav and Sastry, Girish and Askell, Amanda and others},
  journal={Advances in neural information processing systems},
  volume={33},
  pages={1877--1901},
  year={2020}
}

@inproceedings{tang2024once,
  title={Once and for all: Universal transferable adversarial perturbation against deep hashing-based facial image retrieval},
  author={Tang, Long and Ye, Dengpan and Lv, Yunna and Chen, Chuanxi and Zhang, Yunming},
  booktitle={Proceedings of the AAAI Conference on Artificial Intelligence},
  volume={38},
  number={6},
  pages={5136--5144},
  year={2024}
}

@article{zeng2022deep,
  title={Deep generative molecular design reshapes drug discovery},
  author={Zeng, Xiangxiang and Wang, Fei and Luo, Yuan and Kang, Seung-gu and Tang, Jian and Lightstone, Felice C and Fang, Evandro F and Cornell, Wendy and Nussinov, Ruth and Cheng, Feixiong},
  journal={Cell Reports Medicine},
  volume={3},
  number={12},
  year={2022},
  publisher={Elsevier}
}

@inproceedings{ICLR2024_081b0806,
 author = {Podell, Dustin and English, Zion and Lacey, Kyle and Blattmann, Andreas and Dockhorn, Tim and M\"{u}ller, Jonas and Penna, Joe and Rombach, Robin},
 booktitle = {International Conference on Representation Learning},
 editor = {B. Kim and Y. Yue and S. Chaudhuri and K. Fragkiadaki and M. Khan and Y. Sun},
 pages = {1862--1874},
 title = {SDXL: Improving Latent Diffusion Models for High-Resolution Image Synthesis},
 url = {https://proceedings.iclr.cc/paper_files/paper/2024/file/081b08068e4733ae3e7ad019fe8d172f-Paper-Conference.pdf},
 volume = {2024},
 year = {2024}
}

@article{lai2025deep,
  title={Deep Generative Models for Therapeutic Peptide Discovery: A Comprehensive Review},
  author={Lai, Leshan and Liu, Yuansheng and Song, Bosheng and Li, Keqin and Zeng, Xiangxiang},
  journal={ACM Computing Surveys},
  year={2025},
  publisher={ACM New York, NY}
}

@inproceedings{lin2014microsoft,
  title={Microsoft coco: Common objects in context},
  author={Lin, Tsung-Yi and Maire, Michael and Belongie, Serge and Hays, James and Perona, Pietro and Ramanan, Deva and Doll{\'a}r, Piotr and Zitnick, C Lawrence},
  booktitle={Computer vision--ECCV 2014: 13th European conference, zurich, Switzerland, September 6-12, 2014, proceedings, part v 13},
  pages={740--755},
  year={2014},
  organization={Springer}
}

@inproceedings{liu2015deep,
  title={Deep learning face attributes in the wild},
  author={Liu, Ziwei and Luo, Ping and Wang, Xiaogang and Tang, Xiaoou},
  booktitle={Proceedings of the IEEE international conference on computer vision},
  pages={3730--3738},
  year={2015}
}

@article{russakovsky2015imagenet,
  title={Imagenet large scale visual recognition challenge},
  author={Russakovsky, Olga and Deng, Jia and Su, Hao and Krause, Jonathan and Satheesh, Sanjeev and Ma, Sean and Huang, Zhiheng and Karpathy, Andrej and Khosla, Aditya and Bernstein, Michael and others},
  journal={International journal of computer vision},
  volume={115},
  pages={211--252},
  year={2015},
  publisher={Springer}
}

@article{ye2017deep,
  title={Deep learning hierarchical representations for image steganalysis},
  author={Ye, Jian and Ni, Jiangqun and Yi, Yang},
  journal={IEEE Transactions on Information Forensics and Security},
  volume={12},
  number={11},
  pages={2545--2557},
  year={2017},
  publisher={IEEE}
}

@article{you2020siamese,
  title={A Siamese CNN for image steganalysis},
  author={You, Weike and Zhang, Hong and Zhao, Xianfeng},
  journal={IEEE Transactions on Information Forensics and Security},
  volume={16},
  pages={291--306},
  year={2020},
  publisher={IEEE}
}

@article{guan2022deepmih,
  title={DeepMIH: Deep invertible network for multiple image hiding},
  author={Guan, Zhenyu and Jing, Junpeng and Deng, Xin and Xu, Mai and Jiang, Lai and Zhang, Zhou and Li, Yipeng},
  journal={IEEE Transactions on Pattern Analysis and Machine Intelligence},
  volume={45},
  number={1},
  pages={372--390},
  year={2022},
  publisher={IEEE}
}

@article{boroumand2018deep,
  title={Deep residual network for steganalysis of digital images},
  author={Boroumand, Mehdi and Chen, Mo and Fridrich, Jessica},
  journal={IEEE Transactions on Information Forensics and Security},
  volume={14},
  number={5},
  pages={1181--1193},
  year={2018},
  publisher={IEEE}
}

@article{pietikainen2010local,
  title={Local binary patterns},
  author={Pietik{\"a}inen, Matti},
  journal={Scholarpedia},
  volume={5},
  number={3},
  pages={9775},
  year={2010}
}

@article{zeng2023robust,
  title={Robust steganography for high quality images},
  author={Zeng, Kai and Chen, Kejiang and Zhang, Weiming and Wang, Yaofei and Yu, Nenghai},
  journal={IEEE Transactions on Circuits and Systems for Video Technology},
  volume={33},
  number={9},
  pages={4893--4906},
  year={2023},
  publisher={IEEE}
}

@article{ojala2002multiresolution,
  title={Multiresolution gray-scale and rotation invariant texture classification with local binary patterns},
  author={Ojala, Timo and Pietikainen, Matti and Maenpaa, Topi},
  journal={IEEE Transactions on pattern analysis and machine intelligence},
  volume={24},
  number={7},
  pages={971--987},
  year={2002},
  publisher={IEEE}
}

@article{tao2018towards,
  title={Towards robust image steganography},
  author={Tao, Jinyuan and Li, Sheng and Zhang, Xinpeng and Wang, Zichi},
  journal={IEEE Transactions on Circuits and Systems for Video Technology},
  volume={29},
  number={2},
  pages={594--600},
  year={2018},
  publisher={IEEE}
}

@article{meng2025robust,
  author    = {Laijin Meng and Xinghao Jiang and Qiang Xu and Tanfeng Sun},
  title     = {A Robust Coverless Video Steganography Based on Two-level DCT Features Against Video Attacks},
  journal   = {IEEE Transactions on Multimedia},
  year      = {2025},
  publisher = {IEEE}
}

@inproceedings{yang2023autostegafont,
  title={AutoStegaFont: Synthesizing vector fonts for hiding information in documents},
  author={Yang, Xi and Zhang, Jie and Fang, Han and Liu, Chang and Ma, Zehua and Zhang, Weiming and Yu, Nenghai},
  booktitle={Proceedings of the AAAI Conference on Artificial Intelligence},
  volume={37},
  number={3},
  pages={3198--3205},
  year={2023}
}

@inproceedings{xue2025physical,
  title={Physical Marker: Revealing Invisible Hyperlinks Hidden in Printed Trademarks},
  author={Xue, Yuliang and Tan, Lei and Li, Guobiao and Qian, Zhenxing and Li, Sheng and Zhang, Xinpeng},
  booktitle={Proceedings of the AAAI Conference on Artificial Intelligence},
  volume={39},
  number={9},
  pages={9068--9075},
  year={2025}
}

@inproceedings{ji2025speech,
  title={Speech watermarking with discrete intermediate representations},
  author={Ji, Shengpeng and Jiang, Ziyue and Zuo, Jialong and Fang, Minghui and Chen, Yifu and Jin, Tao and Zhao, Zhou},
  booktitle={Proceedings of the AAAI Conference on Artificial Intelligence},
  volume={39},
  number={23},
  pages={24239--24247},
  year={2025}
}

@article{CHENG2025127598,
title = {Robust steganography with boundary-preserving overflow alleviation and adaptive error correction},
journal = {Expert Systems with Applications},
pages = {127598},
year = {2025},
issn = {0957-4174},
author = {Yu Cheng and Zhenlin Luo and Zhaoxia Yin}
}

\appendix                           
\setcounter{secnumdepth}{2}

\section*{Appendix}                  
\addcontentsline{toc}{section}{Appendix}

\section{Ablation Experiment}       
\label{ablation experiment all}


\subsection{Ablation Experiment 1: Texture-aware Localization}
\label{ablation1}
\begin{figure*}[t]          
  \centering
  \subfigure[]{%
    \includegraphics[width=0.28\linewidth]{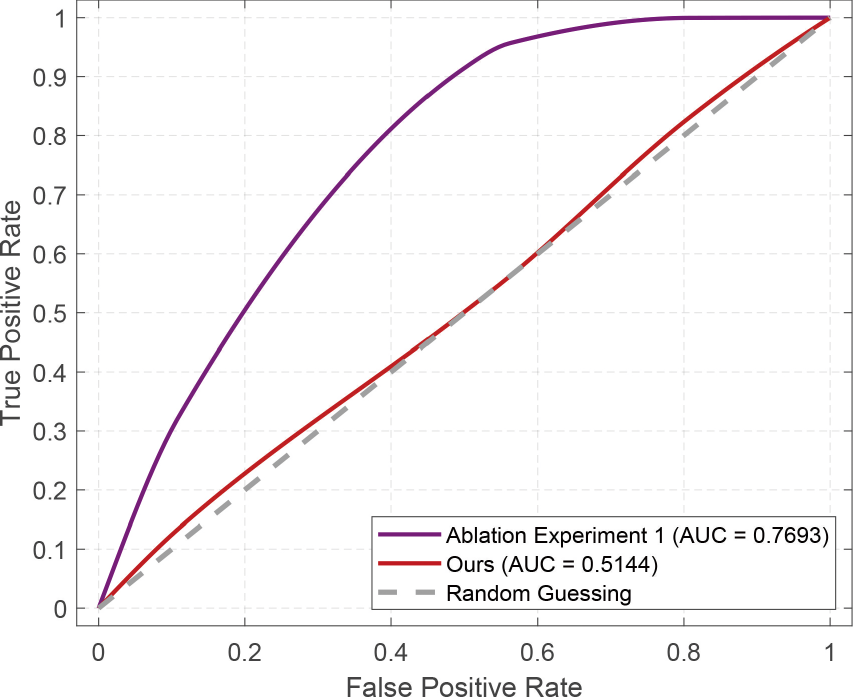}
    \label{fig:fenxi_xiaorong_a}}
  \hfill%
  \subfigure[]{%
    \includegraphics[width=0.28\linewidth]{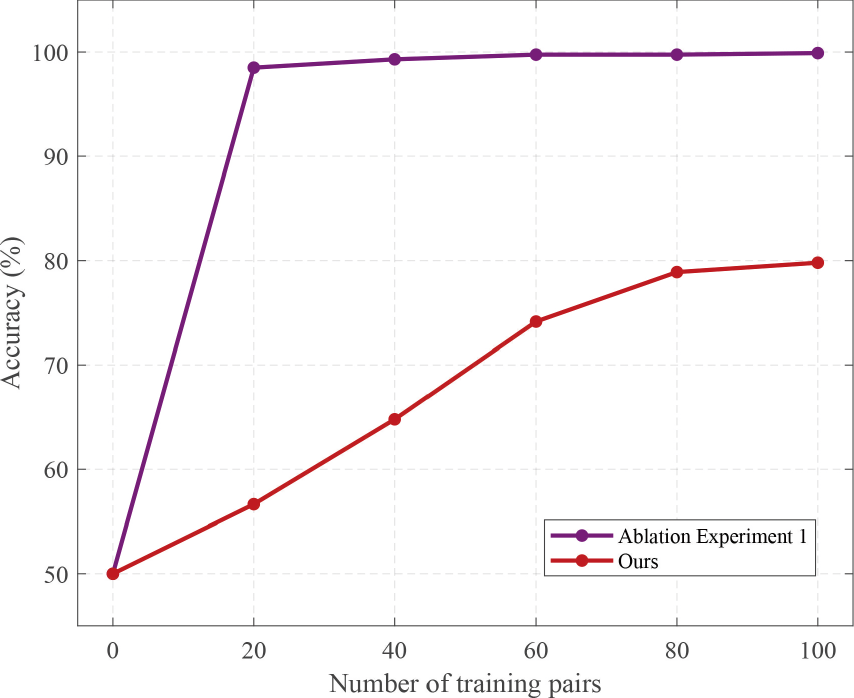}
    \label{fig:fenxi_xiaorong_b}}
  \hfill%
  \subfigure[]{%
    \includegraphics[width=0.28\linewidth]{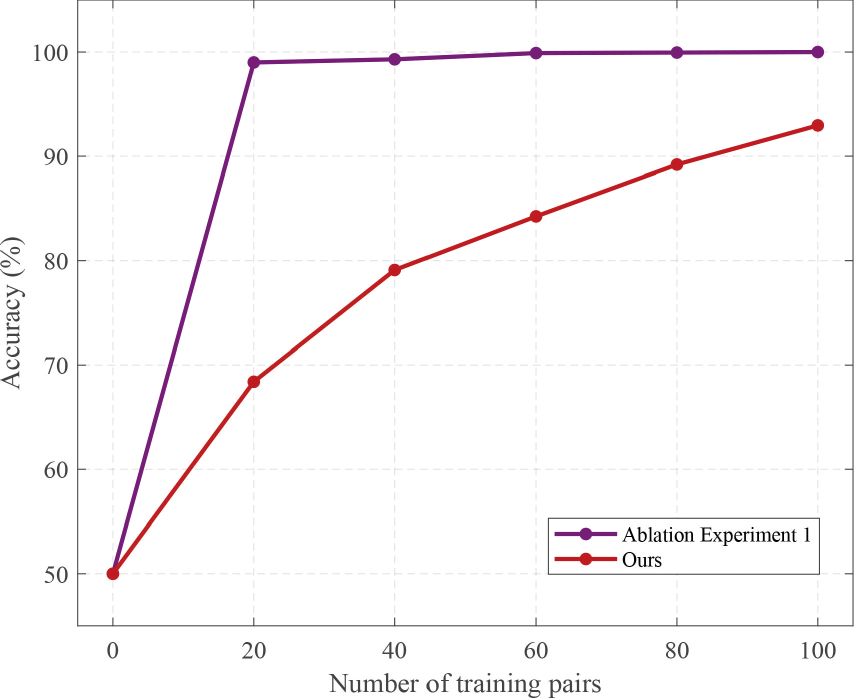}
    \label{fig:fenxi_xiaorong_c}}

  \subfigure[]{%
    \includegraphics[width=0.28\linewidth]{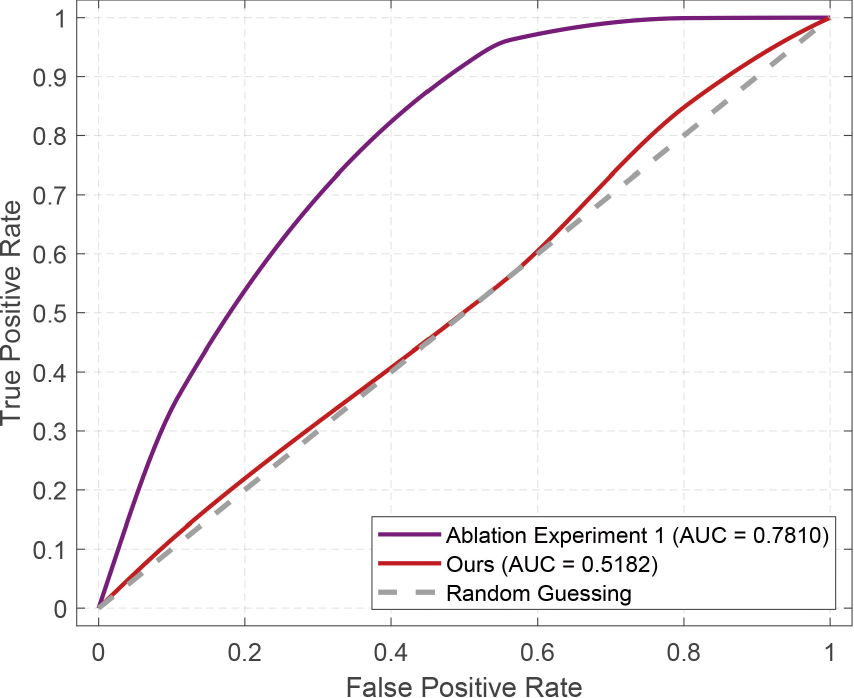}
    \label{fig:fenxi_xiaorong_a2}}
  \hfill%
  \subfigure[]{%
    \includegraphics[width=0.28\linewidth]{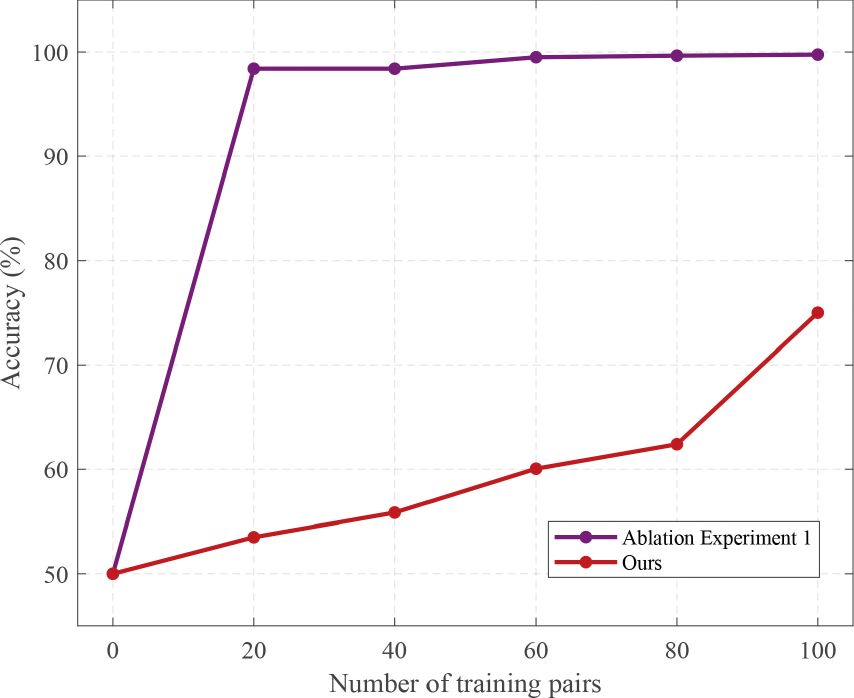}
    \label{fig:fenxi_xiaorong_b2}}
  \hfill%
  \subfigure[]{%
    \includegraphics[width=0.28\linewidth]{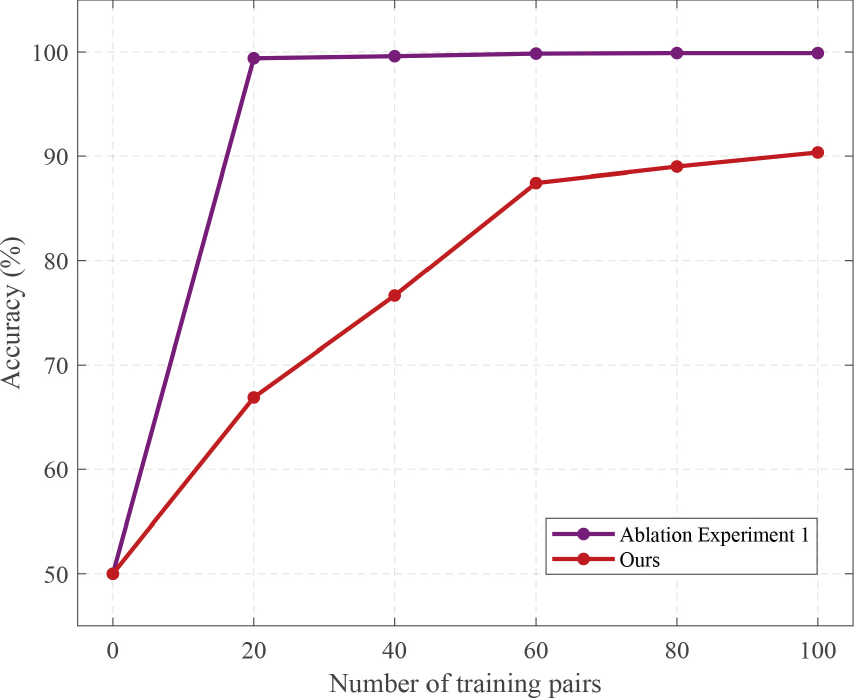}
    \label{fig:fenxi_xiaorong_c2}}

  \caption{Ablation experiment 1: Anti-steganalysis performance of stego images generated by different methods against (a), (d) StegExpose, (b), (e) YeNet, and (c), (f) SiaStegNet.  
  The top row corresponds to low embedding capacity (1.5\,bpp), whereas the bottom row corresponds to high embedding capacity (6\,bpp).}
  \label{fig:fenxi_xiaorong}
\end{figure*}

In this section, we conduct an ablation study, referred to as Ablation Experiment 1, in which the full RFNNS method is compared with a method that excludes the texture‑aware localization technique. The experimental settings are the same as those in Section \ref{settings}. As shown in Table \ref{tab:stego_all_xiaorong} and Table \ref{tab:recovered_secret_all_xiaorong}, when the texture‑aware localization technique is not used, the visual quality of the secret images extracted remains largely unchanged, while the quality of the stego images decreases significantly. In addition, as shown in Fig. \ref{fig:fenxi_xiaorong}, the anti-steganalysis performance of the generated stego images is considerably lower than that of the RFNNS method. Specifically, omitting the texture‑aware localization technique leads to a pronounced decrease in the quality of the stego image with respect to imperceptibility, accompanied by a substantial reduction in anti‑steganalysis performance. This is due to the texture-aware localization technique, which divides the cover image into blocks, assesses their texture complexity, and selects appropriate regions for perturbation to maintain visual quality. Although employing this technique results in a slight decrease in robustness, its performance gap relative to the ablation method remains minimal. Given that security is the most important guarantee for covert communication, we consider the minor trade‑off in robustness to be entirely acceptable.

\subsection{Ablation Experiment 2: Robust Steganographic Perturbation Generation}
\label{ablation2}
In this section, we conduct an ablation study, referred to as Ablation Experiment 2, in which the full RFNNS method is compared with a method that excludes the robust steganographic perturbation generation (RSPG) strategy. The experimental settings are the same as those in Section \ref{settings}. As shown in Table \ref{tab:stego_all_xiaorong2} and Table \ref{tab:recovered_secret_all_xiaorong2}, the visual quality of the stego images is approximately comparable to that of RFNNS, while the quality of the secret images extracted deteriorates significantly. Specifically, the RSPG strategy progressively reduces the discrepancy between the recovered and original secret images during the iterative optimization of the perturbations. In each optimization iteration, it introduces simulated image attack scenarios, which substantially enhances the robustness of the resulting stego images. Although employing this strategy entails a minor decline in stego image quality, it guarantees a commendable level of robustness. Under the low embedding capacity condition and in all contrast adjustments evaluated, the average PSNR of the recovered secret images increases by 6 dB, while the LPIPS value is only 20\% of that achieved by the comparison methods. Under the high embedding capacity, and across all evaluated Gaussian noise attacks, the SSIM of the recovered secret images increases by approximately 35\%. Through this ablation experiment, the remarkable enhancement of robustness brought about by the proposed RSPG strategy is validated.

\section{Robustness and Generalization Evaluation}
\label{robust}
In this section, we evaluate the robustness and generalization ability of RFNNS against leading FNNS schemes under various common image attacks as well as previously unknown attacks. The experimental settings are the same as those in Section \ref{settings}. As presented in Table \ref{tab9} and Table \ref{tab10}, RFNNS consistently outperforms SOTA methods across all evaluation metrics. Specifically, regarding robustness, the LPIPS of secret images recovered by RFNNS under image rotation and scaling attacks are 16\% of those achieved by SOTA methods. As shown in Table \ref{tab5}, regarding generalization performance, RFNNS improves the SSIM of recovered secret images by approximately 17\%. These experimental results clearly demonstrate that RFNNS exhibits superior robustness and strong generalization capabilities, making it highly valuable for practical covert communication scenarios.

\begin{table}[t]
  \centering
  \begin{small} 
  \begin{tabular}{@{}l@{\hspace{3pt}}l@{\hspace{3pt}}
      *{6}{>{\centering\arraybackslash}c@{\hspace{2pt}}}@{}}
    \toprule
    \multirow{2}{*}{Type} & \multirow{2}{*}{Capacity} &
      \multicolumn{3}{c}{Li et al.} &
      \multicolumn{3}{c}{Ours} \\[-1pt]
    \cmidrule(r){3-5}\cmidrule(l){6-8}
      & & PSNR$\uparrow$ & SSIM$\uparrow$ & LPIPS$\downarrow$
        & PSNR$\uparrow$ & SSIM$\uparrow$ & LPIPS$\downarrow$ \\
    \midrule
    Type 1 & 1.5 bpp & 25.27 & 0.809 & 0.123 & \textbf{29.06} & \textbf{0.873} & \textbf{0.075}\\
    Type 2 & 1.5 bpp & 28.68 & 0.868 & 0.061 & \textbf{31.76} & \textbf{0.937} & \textbf{0.025}\\
    Type 3 & 1.5 bpp & 23.56 & 0.785 & 0.122 & \textbf{28.54} & \textbf{0.853} & \textbf{0.090}\\
    Type 4 & 1.5 bpp & 18.20 & 0.646 & 0.329 & \textbf{28.72} & \textbf{0.847} & \textbf{0.099}\\
    \midrule
    Type 1 & 6 bpp   & 18.89 & 0.646 & 0.317 & \textbf{26.36} & \textbf{0.790} & \textbf{0.164}\\
    Type 2 & 6 bpp   & 25.27 & 0.808 & 0.123 & \textbf{29.34} & \textbf{0.881} & \textbf{0.054}\\
    Type 3 & 6 bpp   & 16.10 & 0.551 & 0.412 & \textbf{25.76} & \textbf{0.767} & \textbf{0.173}\\
    Type 4 & 6 bpp   & 18.17 & 0.645 & 0.328 & \textbf{25.83} & \textbf{0.755} & \textbf{0.206}\\
    \bottomrule
  \end{tabular}
  \end{small}
  \caption{Recovered secret image quality under different embedding capacities and unknown attack settings. Type 1 applies JPEG compression and Gaussian noise as surrogate attacks, whereas contrast adjustment is used as the actual attack. Type 2 combines Gaussian noise with contrast adjustment during simulation, while Gaussian blurring serves as the real attack. Type 3 simulates JPEG compression, image scaling, and Gaussian noise, with image rotation acting as the true attack. Type 4 employs JPEG compression, contrast adjustment, and Gaussian noise as surrogates, and likewise evaluates robustness against image rotation as the actual attack. (↑ higher is better; ↓ lower is better).}
  \label{tab5}
\end{table}

\section{Anti-Steganalysis Performance Evaluation}
\label{anti-steganalysis performance}
\begin{figure*}[t]
  \centering
  \subfigure[]{%
    \includegraphics[width=0.28\linewidth]{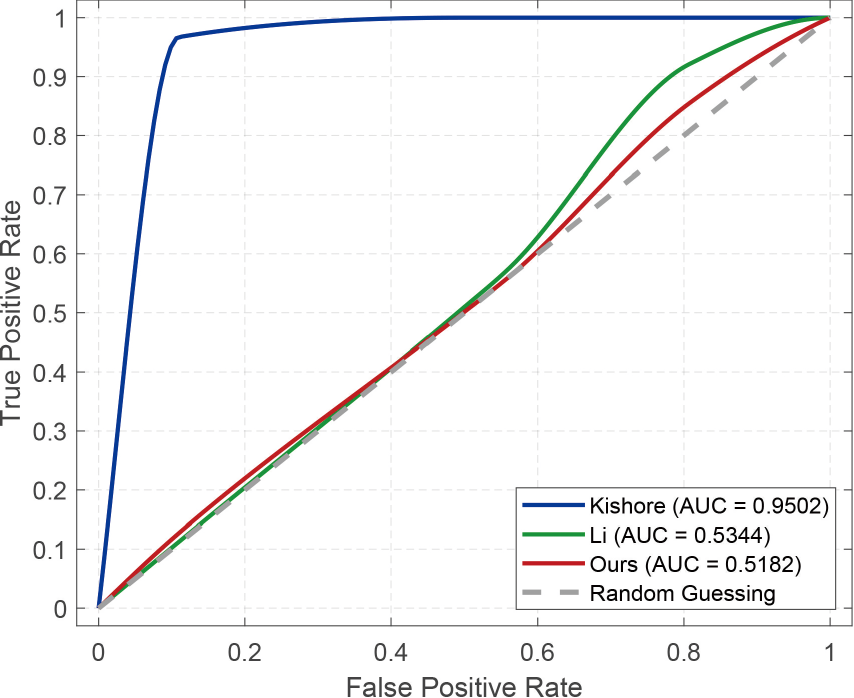}
    \label{fig:fenxi_high_a}}
  \hfill
  \subfigure[]{%
    \includegraphics[width=0.28\linewidth]{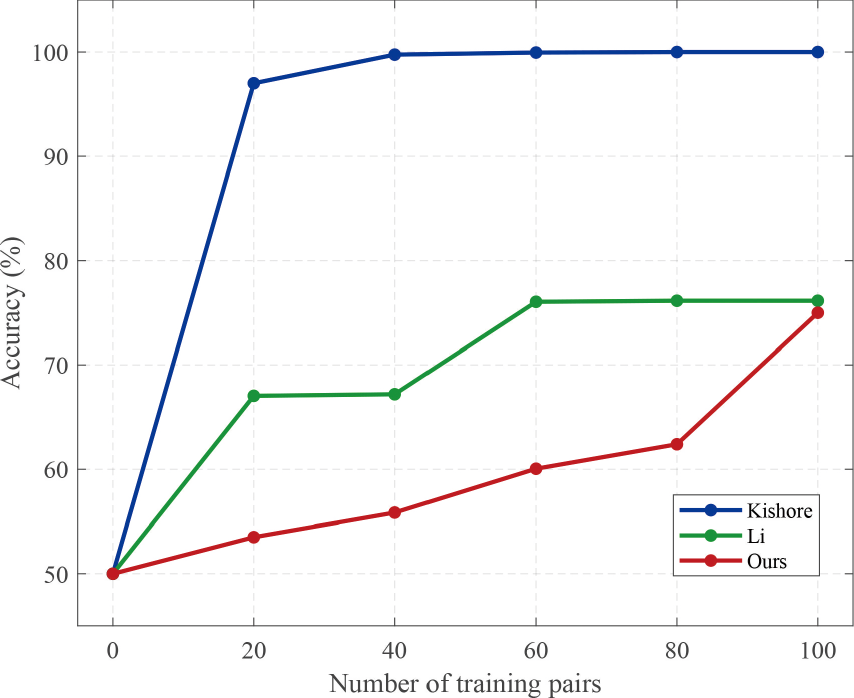}
    \label{fig:fenxi_high_b}}
  \hfill
  \subfigure[]{%
    \includegraphics[width=0.28\linewidth]{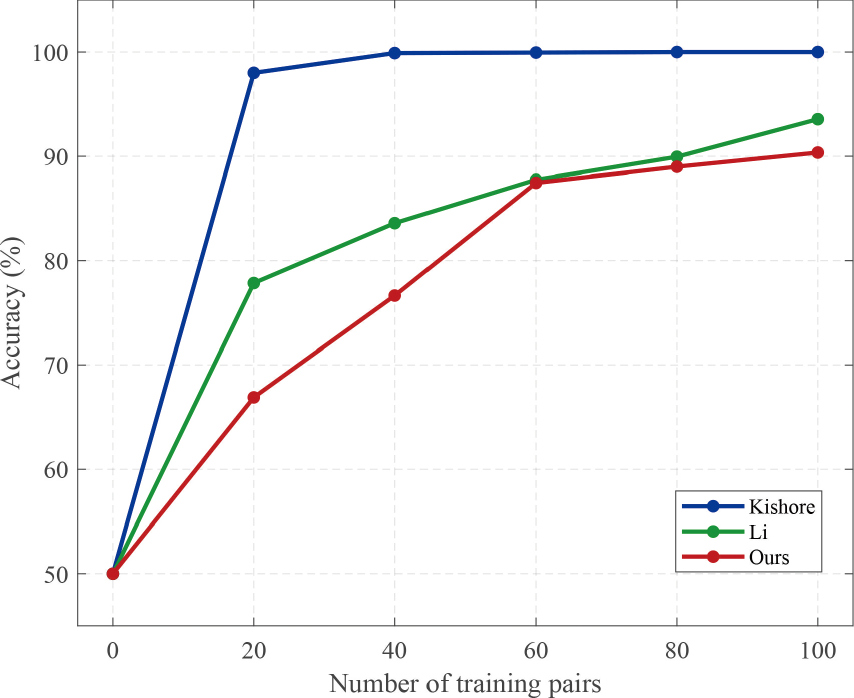}
    \label{fig:fenxi_high_c}}

  \caption{Anti-steganalysis performance at the high embedding
  capacity (6 bpp): (a) StegExpose, (b) YeNet, and (c) SiaStegNet.}
  \label{fig:fenxi_high}
\end{figure*}

\begin{figure*}[t]          
  \centering
  \subfigure[]{%
    \includegraphics[width=0.45\linewidth]{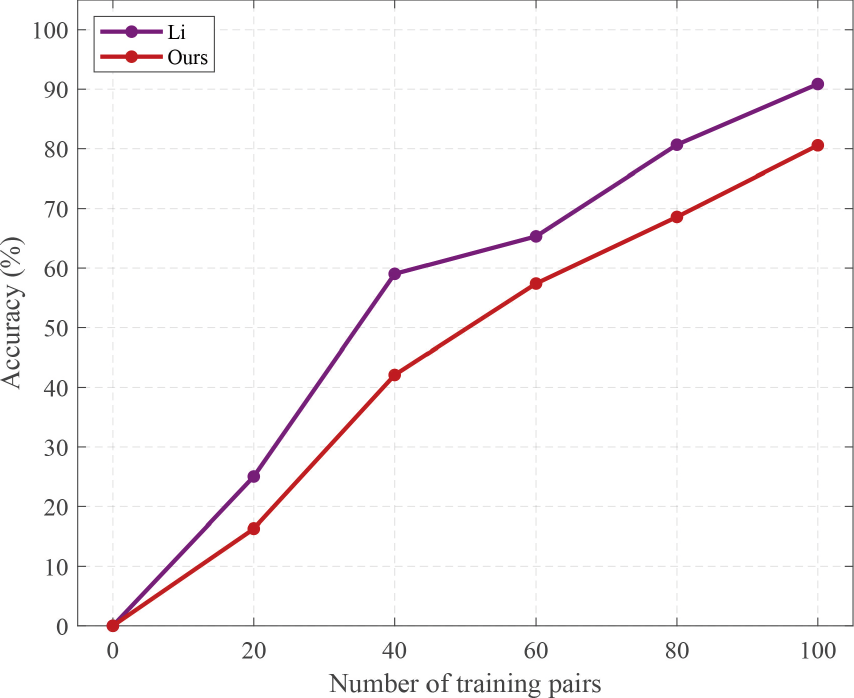}
    \label{fig:fenxi_xiaorong_a}}
  \hfill
  \subfigure[]{%
    \includegraphics[width=0.45\linewidth]{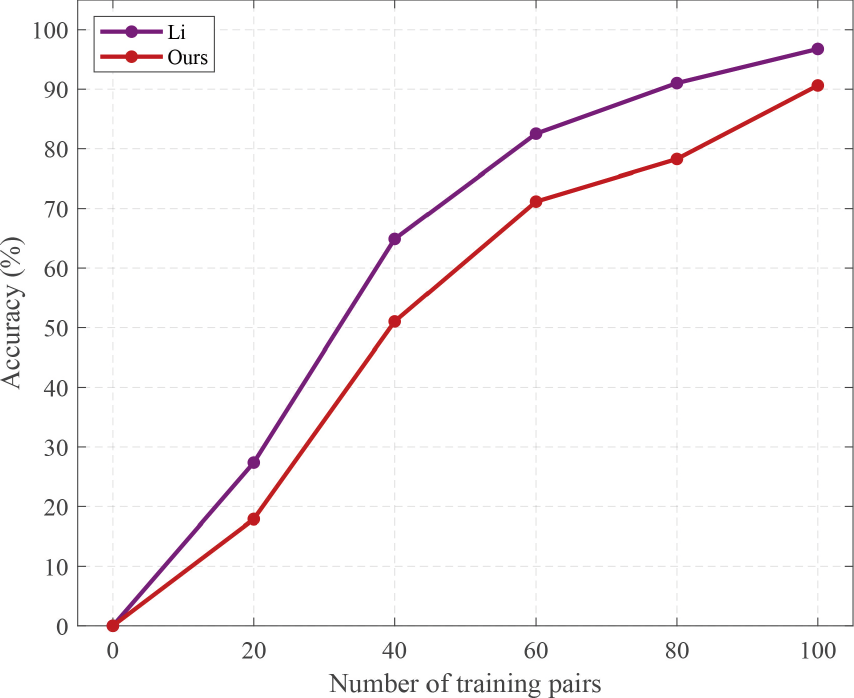}
    \label{fig:fenxi_xiaorong_b}}

  \caption{Anti-Steganalysis performance evaluation with efficientNet-B2 (a) 1.5\,bpp (low) and (b) 6\,bpp (high).}
  \label{fig:new_yinxiefenxi}
\end{figure*}

Due to space constraints, the anti-steganalysis performance at a high embedding capacity (6 bpp), evaluated using StegExpose, YeNet, and SiaStegNet, is presented in this section. The experimental settings are the same as those in Section \ref{settings}. The detailed results are summarized in Fig. \ref{fig:fenxi_high}. Specifically, with 100 training pairs, the detection accuracy of SRNet is limited to 75\%, whereas that of SiaStegNet reaches 90.35\%. To thoroughly evaluate the anti-steganalysis performance of stego images generated by RFNNS, we incorporated the EfficientNet-B2 steganalyzer developed by Fridrich’s group. Specifically, this steganalytic network was trained using corresponding cover-stego image pairs from the training set, with the number of training samples progressively increased during the process. As shown in Fig. \ref{fig:new_yinxiefenxi}, RFNNS consistently achieves lower detection accuracy compared to SOTA FNNS methods, confirming its strong anti-steganalysis performance. Additionally, we applied several widely recognized deep-learning-based steganalyzers, including YeNet, SiaStegNet, and SRNet, in both the perturbation generation and evaluation phases. The overall results indicate that the proposed RFNNS method significantly surpasses existing approaches in terms of anti-steganalysis performance.

\section{Payload}
\label{payload}
In this section, we comprehensively evaluate the performance of RFNNS under different embedding capacities. The experimental settings are the same as those in Section \ref{settings}. As shown in Table \ref{different payload}, RFNNS maintains outstanding performance even at higher embedding capacities. Specifically, at 24 bpp, the method achieves a secret image SSIM score of approximately 0.805, demonstrating impressive practicality in scenarios of high payload embedding. These results clearly highlight the superior capability of RFNNS to maintain reliable image recovery even at high embedding rates, thus confirming its practical effectiveness for covert communication.

\begin{table}[H]
  \centering
  \begin{footnotesize} 
  \begin{tabular}{@{}l@{\hspace{3pt}}
                  *{6}{>{\centering\arraybackslash}c@{\hspace{2pt}}}@{}}
    \toprule
    \multirow{2}{*}{Capacity}
      & \multicolumn{3}{c}{Stego image}
      & \multicolumn{3}{c}{Recovered secret image} \\[-1pt]
    \cmidrule(r){2-4}\cmidrule(l){5-7}
      & PSNR$\uparrow$ & SSIM$\uparrow$ & LPIPS$\downarrow$
        & PSNR$\uparrow$ & SSIM$\uparrow$ & LPIPS$\downarrow$ \\
    \midrule
    0.375 bpp & 35.4  & 0.942 & 0.022 & 32.05 & 0.956 & 0.013 \\
    13.5 bpp  & 33.46 & 0.911 & 0.033 & 21.64 & 0.822 & 0.160 \\
    24 bpp    & 32.32 & 0.887 & 0.041 & 20.97 & 0.805 & 0.291 \\
    \bottomrule
  \end{tabular}
  \caption{Stego image and recovered secret image quality under different embedding capacities.
  (↑ higher is better; ↓ lower is better).}
  \label{different payload}
  \end{footnotesize}
\end{table}

\section{Text-to-Image Model Selection}
In this section, we examine how the choice of text-to-image (T2I) model and the texture complexity of cover images affect RFNNS. Under identical settings, covers generated by SDXL, Stable Cascade, and LDM yield performance variation within ±5\%, indicating that RFNNS is largely model-independent. Stable Diffusion is therefore used as the representative baseline. 

\section{Low-Texture Performance and Texture Metric Choice}
On low-texture covers, the recovered secret remains essentially unchanged, while stego quality shows a slight drop as RFNNS increases perturbations in smooth regions to meet capacity targets; this can be mitigated by dynamically adjusting the threshold. We compute texture complexity using Local Binary Patterns (LBP); this is not our innovation but a standard instantiation of the broader practice of filtering by texture complexity to reduce the perturbation scale.

\section{Computational Efficiency and Hyperparameter Settings}
In this section, we discuss the computational efficiency and hyperparameter settings of RFNNS. RFNNS requires only a shared key and a prompt, avoiding large-model transfer and reducing communication cost. It uses the same 1,500 iterations as the SOTA Cs-FNNS. Embedding takes 0.1-3 min on one RTX 4090 without training, whereas HiNet needs over 120 hours of training on 8 RTX 4090 GPUs, showing a clear cost advantage. The choice of parameter $Y$ is guided by parameter-selection experiments; at this value, it shortens the iteration time while preserving stego image quality.

\begin{table*}[htbp]
  \centering
  \begin{small} 
  \begin{tabular}{@{}l@{\hspace{4pt}}l@{\hspace{4pt}}l
      @{\hspace{6pt}}>{\centering\arraybackslash}p{1.95cm}@{\hspace{2pt}}
      >{\centering\arraybackslash}p{1.95cm}@{\hspace{2pt}}
      >{\centering\arraybackslash}p{1.95cm}@{\hspace{6pt}}
      >{\centering\arraybackslash}p{1.45cm}@{\hspace{2pt}}
      >{\centering\arraybackslash}p{1.45cm}@{\hspace{2pt}}
      >{\centering\arraybackslash}p{1.45cm}@{}}
    \toprule[0.8pt]
    \multirow{2}{*}{Capacity} & \multirow{2}{*}{Attack} & \multirow{2}{*}{Factor} &
      \multicolumn{3}{c}{RFNNS without Texture-Aware Loc.} &
      \multicolumn{3}{c}{RFNNS} \\[-1pt]
    \cmidrule(lr){4-6}\cmidrule(lr){7-9}
     & & & PSNR$\uparrow$ & SSIM$\uparrow$ & LPIPS$\downarrow$
       & PSNR$\uparrow$ & SSIM$\uparrow$ & LPIPS$\downarrow$ \\
    \midrule
    \multirow{10}{*}{1.5 bpp}
      & \multicolumn{2}{l}{No Attack}
          & 30.01 & 0.837 & 0.054
          & \textbf{41.48} & \textbf{0.980} & \textbf{0.003} \\
    \cmidrule{2-9}
      & \multirow{3}{*}{JPEG Compression}
          & \textit{QF}=90
              & 22.18 & 0.527 & 0.229
              & \textbf{25.95} & \textbf{0.717} & \textbf{0.134} \\
      & & \textit{QF}=80
              & 19.75 & 0.425 & 0.328
              & \textbf{25.43} & \textbf{0.703} & \textbf{0.147} \\
      & & \textit{QF}=70
              & 18.14 & 0.360 & 0.411
              & \textbf{22.51} & \textbf{0.608} & \textbf{0.223} \\
    \cmidrule{2-9}
      & \multirow{3}{*}{Gaussian Noise}
          & $\rho$=0.01
              & 29.66 & 0.815 & 0.067
              & \textbf{32.33} & \textbf{0.880} & \textbf{0.046} \\
      & & $\rho$=0.04
              & 24.40 & 0.632 & 0.169
              & \textbf{28.66} & \textbf{0.800} & \textbf{0.089} \\
      & & $\rho$=0.07
              & 22.38 & 0.552 & 0.232
              & \textbf{26.72} & \textbf{0.748} & \textbf{0.124} \\
    \cmidrule{2-9}
      & \multirow{3}{*}{Contrast Adjustment}
          & $\eta$=0.9
              & 30.03 & 0.836 & 0.054
              & \textbf{33.46} & \textbf{0.913} & \textbf{0.041} \\
      & & $\eta$=0.8
              & 30.02 & 0.834 & 0.055
              & \textbf{32.98} & \textbf{0.899} & \textbf{0.043} \\
      & & $\eta$=0.7
              & 29.95 & 0.832 & 0.055
              & \textbf{32.60} & \textbf{0.889} & \textbf{0.047} \\
    \midrule
    \multirow{10}{*}{6 bpp}
      & \multicolumn{2}{l}{No Attack}
          & 30.01 & 0.835 & 0.041
          & \textbf{42.95} & \textbf{0.984} & \textbf{0.003} \\
    \cmidrule{2-9}
      & \multirow{3}{*}{JPEG Compression}
          & \textit{QF}=90
              & 17.23 & 0.327 & 0.357
              & \textbf{22.62} & \textbf{0.583} & \textbf{0.218} \\
      & & \textit{QF}=80
              & 16.65 & 0.304 & 0.410
              & \textbf{21.58} & \textbf{0.565} & \textbf{0.222} \\
      & & QF=70
              & 16.36 & 0.293 & 0.434
              & \textbf{19.81} & \textbf{0.522} & \textbf{0.292} \\
    \cmidrule{2-9}
      & \multirow{3}{*}{Gaussian Noise}
          & $\rho$=0.01
              & 27.48 & 0.743 & 0.085
              & \textbf{31.62} & \textbf{0.864} & \textbf{0.048} \\
      & & $\rho$=0.04
              & 22.34 & 0.550 & 0.199
              & \textbf{28.51} & \textbf{0.786} & \textbf{0.087} \\
      & & $\rho$=0.07
              & 20.53 & 0.477 & 0.251
              & \textbf{26.19} & \textbf{0.738} & \textbf{0.130} \\
    \cmidrule{2-9}
      & \multirow{3}{*}{Contrast Adjustment}
          & $\eta$=0.9
              & 29.90 & 0.823 & 0.044
              & \textbf{32.73} & \textbf{0.908} & \textbf{0.043} \\
      & & $\eta$=0.8
              & 28.65 & 0.787 & \textbf{0.056}
              & \textbf{30.72} & \textbf{0.845} & 0.059 \\
      & & $\eta$=0.7
              & 25.79 & 0.697 & 0.094
              & \textbf{28.15} & \textbf{0.784} & \textbf{0.093} \\
    \bottomrule[1.2pt]
  \end{tabular}
  \end{small}
  \caption{Ablation Experiment 1: Stego image quality under different embedding capacities and attack conditions}
  \label{tab:stego_all_xiaorong}
\end{table*}

\begin{table*}[hb]
  \centering
  \begin{small} 
  \begin{tabular}{@{}l@{\hspace{4pt}}l@{\hspace{4pt}}l
      @{\hspace{6pt}}>{\centering\arraybackslash}p{1.95cm}@{\hspace{2pt}}
      >{\centering\arraybackslash}p{1.95cm}@{\hspace{2pt}}
      >{\centering\arraybackslash}p{1.95cm}@{\hspace{6pt}}
      >{\centering\arraybackslash}p{1.45cm}@{\hspace{2pt}}
      >{\centering\arraybackslash}p{1.45cm}@{\hspace{2pt}}
      >{\centering\arraybackslash}p{1.45cm}@{}}
    \toprule[0.8pt]
    \multirow{2}{*}{Capacity} & \multirow{2}{*}{Attack} & \multirow{2}{*}{Factor} &
      \multicolumn{3}{c}{RFNNS without Texture-Aware Localization} &
      \multicolumn{3}{c}{RFNNS} \\[-1pt]
    \cmidrule(lr){4-6}\cmidrule(lr){7-9}
     & & & PSNR$\uparrow$ & SSIM$\uparrow$ & LPIPS$\downarrow$
       & PSNR$\uparrow$ & SSIM$\uparrow$ & LPIPS$\downarrow$ \\
    \midrule
    \multirow{10}{*}{1.5 bpp}
      & \multicolumn{2}{l}{No Attack}
          & \textbf{39.56} & \textbf{0.980} & \textbf{0.004}
          & 34.14 & 0.943 & 0.017 \\
    \cmidrule{2-9}
      & \multirow{3}{*}{JPEG Compression}
          & \textit{QF}=90
              & \textbf{31.52} & \textbf{0.906} & \textbf{0.037}
              & 29.28 & 0.861 & 0.070 \\
      & & \textit{QF}=80
              & \textbf{29.70} & \textbf{0.880} & \textbf{0.053}
              & 29.27 & 0.858 & 0.072 \\
      & & \textit{QF}=70
              & \textbf{27.93} & \textbf{0.853} & \textbf{0.068}
              & 27.00 & 0.813 & 0.112 \\
    \cmidrule{2-9}
      & \multirow{3}{*}{Gaussian Noise}
          & $\rho$=0.01
              & \textbf{34.34} & \textbf{0.942} & \textbf{0.021}
              & 32.04 & 0.920 & 0.037 \\
      & & $\rho$=0.04
              & \textbf{30.82} & \textbf{0.892} & \textbf{0.048}
              & 27.44 & 0.816 & 0.124 \\
      & & $\rho$=0.07
              & \textbf{29.12} & \textbf{0.860} & \textbf{0.068}
              & 26.08 & 0.756 & 0.169 \\
    \cmidrule{2-9}
      & \multirow{3}{*}{Contrast Adjustment}
          & $\eta$=0.9
              & \textbf{40.17} & \textbf{0.981} & \textbf{0.003}
              & 34.62 & 0.968 & 0.016 \\
      & & $\eta$=0.8
              & \textbf{40.16} & \textbf{0.980} & \textbf{0.003}
              & 34.38 & 0.953 & 0.017 \\
      & & $\eta$=0.7
              & \textbf{40.02} & \textbf{0.977} & \textbf{0.003}
              & 33.68 & 0.950 & 0.019 \\
    \midrule
    \multirow{10}{*}{6 bpp}
      & \multicolumn{2}{l}{No Attack}
          & \textbf{38.56} & \textbf{0.963} & \textbf{0.009}
          & 31.09 & 0.910 & 0.058 \\
    \cmidrule{2-9}
      & \multirow{3}{*}{JPEG Compression}
          & \textit{QF}=90
              & 22.26 & \textbf{0.760} & \textbf{0.184}
              & \textbf{23.60} & 0.720 & 0.253 \\
      & & \textit{QF}=80
              & 20.20 & \textbf{0.706} & \textbf{0.248}
              & \textbf{22.85} & 0.696 & 0.260 \\
      & & \textit{QF}=70
              & 19.00 & \textbf{0.667} & \textbf{0.296}
              & \textbf{19.24} & 0.572 & 0.411 \\
    \cmidrule{2-9}
      & \multirow{3}{*}{Gaussian Noise}
          & $\rho$=0.01
              & \textbf{32.43} & \textbf{0.902} & \textbf{0.051}
              & 30.07 & 0.855 & 0.117 \\
      & & $\rho$=0.04
              & \textbf{27.93} & \textbf{0.827} & \textbf{0.107}
              & 26.94 & 0.751 & 0.203 \\
      & & $\rho$=0.07
              & \textbf{25.00} & \textbf{0.776} & \textbf{0.153}
              & 24.49 & 0.665 & 0.294 \\
    \cmidrule{2-9}
      & \multirow{3}{*}{Contrast Adjustment}
          & $\eta$=0.9
              & \textbf{37.25} & \textbf{0.952} & \textbf{0.011}
              & 32.79 & 0.919 & 0.030 \\
      & & $\eta$=0.8
              & \textbf{33.76} & \textbf{0.933} & \textbf{0.021}
              & 30.67 & 0.898 & 0.043 \\
      & & $\eta$=0.7
              & \textbf{30.53} & \textbf{0.912} & \textbf{0.039}
              & 28.69 & 0.879 & 0.071 \\
    \bottomrule[1.2pt]
  \end{tabular}
  \end{small}
  \caption{Ablation Experiment 1: Recovered secret image quality under different embedding capacities and attack conditions}
  \label{tab:recovered_secret_all_xiaorong}
\end{table*}

\begin{table*}[htbp]
  \centering
  \begin{small} 
  \begin{tabular}{@{}l@{\hspace{4pt}}l@{\hspace{4pt}}l
      @{\hspace{6pt}}>{\centering\arraybackslash}p{1.95cm}@{\hspace{2pt}}
      >{\centering\arraybackslash}p{1.95cm}@{\hspace{2pt}}
      >{\centering\arraybackslash}p{1.95cm}@{\hspace{6pt}}
      >{\centering\arraybackslash}p{1.45cm}@{\hspace{2pt}}
      >{\centering\arraybackslash}p{1.45cm}@{\hspace{2pt}}
      >{\centering\arraybackslash}p{1.45cm}@{}}
    \toprule[0.8pt]
    \multirow{2}{*}{Capacity} & \multirow{2}{*}{Attack} & \multirow{2}{*}{Factor} &
      \multicolumn{3}{c}{RFNNS without the RSPG strategy} &
      \multicolumn{3}{c}{RFNNS} \\[-1pt]
    \cmidrule(lr){4-6}\cmidrule(lr){7-9}
     & & & PSNR$\uparrow$ & SSIM$\uparrow$ & LPIPS$\downarrow$
       & PSNR$\uparrow$ & SSIM$\uparrow$ & LPIPS$\downarrow$ \\
    \midrule
    \multirow{10}{*}{1.5 bpp}
      & \multicolumn{2}{l}{No Attack}
          & \textbf{46.24} & 0.963 & \textbf{0.001}
          & 41.48 & \textbf{0.980} & 0.003 \\
    \cmidrule{2-9}
      & \multirow{3}{*}{JPEG Compression}
          & \textit{QF}=90
              & \textbf{29.48} & \textbf{0.820} & \textbf{0.119}
              & 25.95 & 0.717 & 0.134 \\
      & & \textit{QF}=80
              & \textbf{26.88} & \textbf{0.745} & \textbf{0.185}
              & 25.43 & 0.703 & 0.147 \\
      & & \textit{QF}=70
              & \textbf{25.54} & \textbf{0.702} & 0.236
              & 22.51 & 0.608 & \textbf{0.223} \\
    \cmidrule{2-9}
      & \multirow{3}{*}{Gaussian Noise}
          & $\rho$=0.01
              & \textbf{33.96} & \textbf{0.905} & \textbf{0.033}
              & 32.33 & 0.880 & 0.046 \\
      & & $\rho$=0.04
              & 26.04 & 0.680 & 0.140
              & \textbf{28.66} & \textbf{0.800} & \textbf{0.089} \\
      & & $\rho$=0.07
              & 22.26 & 0.530 & 0.262
              & \textbf{26.72} & \textbf{0.748} & \textbf{0.124} \\
    \cmidrule{2-9}
      & \multirow{3}{*}{Contrast Adjustment}
          & $\eta$=0.9
              & 31.35 & \textbf{0.924} & \textbf{0.020}
              & \textbf{33.46} & 0.913 & 0.041 \\
      & & $\eta$=0.8
              & 27.02 & \textbf{0.916} & \textbf{0.031}
              & \textbf{32.98} & 0.899 & 0.043 \\
      & & $\eta$=0.7
              & 22.92 & 0.866 & 0.066
              & \textbf{32.60} & \textbf{0.889} & \textbf{0.047} \\
    \midrule
    \multirow{10}{*}{6 bpp}
      & \multicolumn{2}{l}{No Attack}
          & \textbf{47.31} & 0.979 & \textbf{0.001}
          & 42.95 & \textbf{0.984} & 0.003 \\
    \cmidrule{2-9}
      & \multirow{3}{*}{JPEG Compression}
          & \textit{QF}=90
              & \textbf{26.20} & \textbf{0.719} & \textbf{0.198}
              & 22.62 & 0.583 & 0.218 \\
      & & \textit{QF}=80
              & \textbf{24.78} & \textbf{0.673} & \textbf{0.247}
              & 21.58 & 0.565 & 0.222 \\
      & & \textit{QF}=70
              & \textbf{24.10} & \textbf{0.650} & \textbf{0.276}
              & 19.81 & 0.522 & 0.292 \\
    \cmidrule{2-9}
      & \multirow{3}{*}{Gaussian Noise}
          & $\rho$=0.01
              & \textbf{32.56} & \textbf{0.879} & \textbf{0.043}
              & 31.62 & 0.864 & 0.048 \\
      & & $\rho$=0.04
              & 25.21 & 0.650 & 0.158
              & \textbf{28.51} & \textbf{0.786} & \textbf{0.087} \\
      & & $\rho$=0.07
              & 21.80 & 0.509 & 0.281
              & \textbf{26.19} & \textbf{0.738} & \textbf{0.130} \\
    \cmidrule{2-9}
      & \multirow{3}{*}{Contrast Adjustment}
          & $\eta$=0.9
              & 31.32 & \textbf{0.924} & \textbf{0.020}
              & \textbf{32.73} & 0.908 & 0.043 \\
      & & $\eta$=0.8
              & 25.63 & 0.852 & \textbf{0.056}
              & \textbf{30.72} & 0.845 & 0.059 \\
      & & $\eta$=0.7
              & 22.41 & \textbf{0.789} & 0.100
              & \textbf{28.15} & 0.784 & \textbf{0.093} \\
    \bottomrule[1.2pt]
  \end{tabular}
  \end{small}
  \caption{Ablation Experiment 2: Stego image quality under different embedding capacities and attack conditions}
  \label{tab:stego_all_xiaorong2}
\end{table*}

\begin{table*}[htbp]
  \centering
  \begin{small} 
  \begin{tabular}{@{}l@{\hspace{4pt}}l@{\hspace{4pt}}l
      @{\hspace{6pt}}>{\centering\arraybackslash}p{1.95cm}@{\hspace{2pt}}
      >{\centering\arraybackslash}p{1.95cm}@{\hspace{2pt}}
      >{\centering\arraybackslash}p{1.95cm}@{\hspace{6pt}}
      >{\centering\arraybackslash}p{1.45cm}@{\hspace{2pt}}
      >{\centering\arraybackslash}p{1.45cm}@{\hspace{2pt}}
      >{\centering\arraybackslash}p{1.45cm}@{}}
    \toprule[0.8pt]
    \multirow{2}{*}{Capacity} & \multirow{2}{*}{Attack} & \multirow{2}{*}{Factor} &
      \multicolumn{3}{c}{RFNNS without the RSPG strategy} &
      \multicolumn{3}{c}{RFNNS} \\[-1pt]
    \cmidrule(lr){4-6}\cmidrule(lr){7-9}
     & & & PSNR$\uparrow$ & SSIM$\uparrow$ & LPIPS$\downarrow$
       & PSNR$\uparrow$ & SSIM$\uparrow$ & LPIPS$\downarrow$ \\
    \midrule
    \multirow{10}{*}{1.5 bpp}
      & \multicolumn{2}{l}{No Attack}
          & 30.77 & 0.892 & 0.037
          & \textbf{34.14} & \textbf{0.943} & \textbf{0.017} \\
    \cmidrule{2-9}
      & \multirow{3}{*}{JPEG Compression}
          & \textit{QF}=90
              & 27.28 & 0.799 & 0.130
              & \textbf{29.28} & \textbf{0.861} & \textbf{0.070} \\
      & & \textit{QF}=80
              & 26.06 & 0.761 & 0.131
              & \textbf{29.27} & \textbf{0.858} & \textbf{0.072} \\
      & & \textit{QF}=70
              & 23.95 & 0.726 & 0.191
              & \textbf{27.00} & \textbf{0.813} & \textbf{0.112} \\
    \cmidrule{2-9}
      & \multirow{3}{*}{Gaussian Noise}
          & $\rho$=0.01
              & 28.91 & 0.843 & 0.094
              & \textbf{32.04} & \textbf{0.920} & \textbf{0.037} \\
      & & $\rho$=0.04
              & 23.53 & 0.661 & 0.272
              & \textbf{27.44} & \textbf{0.816} & \textbf{0.124} \\
      & & $\rho$=0.07
              & 21.10 & 0.570 & 0.378
              & \textbf{26.08} & \textbf{0.756} & \textbf{0.169} \\
    \cmidrule{2-9}
      & \multirow{3}{*}{Contrast Adjustment}
          & $\eta$=0.9
              & 15.44 & 0.412 & 0.606
              & \textbf{34.62} & \textbf{0.968} & \textbf{0.016} \\
      & & $\eta$=0.8
              & 14.48 & 0.357 & 0.621
              & \textbf{34.38} & \textbf{0.953} & \textbf{0.017} \\
      & & $\eta$=0.7
              & 13.86 & 0.331 & 0.636
              & \textbf{33.68} & \textbf{0.950} & \textbf{0.019} \\
    \midrule
    \multirow{10}{*}{6 bpp}
      & \multicolumn{2}{l}{No Attack}
          & 28.97 & 0.837 & 0.107
          & \textbf{31.09} & \textbf{0.910} & \textbf{0.058} \\
    \cmidrule{2-9}
      & \multirow{3}{*}{JPEG Compression}
          & \textit{QF}=90
              & 20.21 & 0.598 & 0.390
              & \textbf{23.60} & \textbf{0.720} & \textbf{0.253} \\
      & & \textit{QF}=80
              & 18.95 & 0.543 & 0.454
              & \textbf{22.85} & \textbf{0.696} & \textbf{0.260} \\
      & & \textit{QF}=70
              & 17.89 & 0.497 & 0.511
              & \textbf{19.24} & 0.572 & 0.411 \\
    \cmidrule{2-9}
      & \multirow{3}{*}{Gaussian Noise}
          & $\rho$=0.01
              & 26.57 & 0.748 & 0.227
              & \textbf{30.07} & \textbf{0.855} & \textbf{0.117} \\
      & & $\rho$=0.04
              & 20.39 & 0.523 & 0.481
              & \textbf{26.94} & \textbf{0.751} & \textbf{0.203} \\
      & & $\rho$=0.07
              & 17.78 & 0.422 & 0.605
              & \textbf{24.49} & \textbf{0.665} & \textbf{0.294} \\
    \cmidrule{2-9}
      & \multirow{3}{*}{Contrast Adjustment}
          & $\eta$=0.9
              & 15.77 & 0.413 & 0.604
              & \textbf{32.79} & 0.919 & 0.030 \\
      & & $\eta$=0.8
              & 14.63 & 0.360 & 0.660
              & \textbf{30.67} & \textbf{0.898} & \textbf{0.043} \\
      & & $\eta$=0.7
              & 13.95 & 0.323 & 0.711
              & \textbf{28.69} & \textbf{0.879} & \textbf{0.071} \\
    \bottomrule[1.2pt]
  \end{tabular}
  \end{small}
  \caption{Ablation Experiment 2: Recovered secret image quality under different embedding capacities and attack conditions}
  \label{tab:recovered_secret_all_xiaorong2}
\end{table*}

\begin{table*}[htbp]
  \centering
  \begin{small}          
  \begin{tabular}{@{}l l l l l l l l l l l l l@{}} 
    \toprule[0.8pt]
    \multirow{2}{*}{Capacity} & \multirow{2}{*}{Attack} & \multirow{2}{*}{Factor} &
      \multicolumn{3}{c}{Kishore et al.} &
      \multicolumn{3}{c}{Li et al.} &
      \multicolumn{3}{c}{Ours} \\[-1pt]   
    \cmidrule(lr){4-6}\cmidrule(lr){7-9}\cmidrule(lr){10-12}
    & & & PSNR$\uparrow$ & SSIM$\uparrow$ & LPIPS$\downarrow$
        & PSNR$\uparrow$ & SSIM$\uparrow$ & LPIPS$\downarrow$
        & PSNR$\uparrow$ & SSIM$\uparrow$ & LPIPS$\downarrow$ \\
    \midrule
    \multirow{8}{*}{1.5 bpp}
      & Image Scaling   & $s$=0.95
          & 14.59 & 0.363 & 0.723
          & 31.29 & \textbf{0.900} & \textbf{0.032}
          & \textbf{32.31} & 0.887 & 0.040 \\
    \cmidrule{2-12}
      & Image Rotation  & $\sigma$=0.1
          & 19.94 & 0.621 & 0.346
          & 28.19 & 0.804 & 0.058
          & \textbf{32.29} & \textbf{0.888} & \textbf{0.037} \\
    \cmidrule{2-12}
      & \multirow{2}{*}{JPEG Compression}
          & \textit{QF}=90
              & 14.00 & 0.213 & 1.051
              & 25.26 & 0.659 & 0.244
              & \textbf{25.95} & \textbf{0.717} & \textbf{0.134} \\
      & & \textit{QF}=70
              & 13.99 & 0.212 & 1.059
              & 22.24 & 0.539 & 0.399
              & \textbf{22.51} & \textbf{0.608} & \textbf{0.223} \\
    \cmidrule{2-12}
      & \multirow{2}{*}{Gaussian Noise}
          & $\rho$=0.01
              & 14.31 & 0.200 & 0.881
              & 30.17 & 0.828 & 0.062
              & \textbf{32.33} & \textbf{0.880} & \textbf{0.046} \\
      & & $\rho$=0.04
              & 13.95 & 0.194 & 0.900
              & 23.78 & 0.598 & 0.197
              & \textbf{28.66} & \textbf{0.800} & \textbf{0.089} \\
    \cmidrule{2-12}
      & \multirow{2}{*}{Contrast Adj.}
          & $\eta$=0.9
              & 13.33 & 0.349 & 0.693
              & \textbf{34.36} & \textbf{0.964} & \textbf{0.008}
              & 33.46 & 0.913 & 0.041 \\
      & & $\eta$=0.8
              & 13.08 & 0.389 & 0.643
              & 28.88 & \textbf{0.934} & \textbf{0.015}
              & \textbf{32.98} & 0.899 & 0.043 \\
    \midrule
    \multirow{8}{*}{6 bpp}
      & Image Scaling   & $s$=0.95
          & 14.59 & 0.363 & 0.723
          & 28.22 & 0.816 & 0.092
          & \textbf{29.75} & \textbf{0.831} & \textbf{0.060} \\
    \cmidrule{2-12}
      & Image Rotation  & $\sigma$=0.1
          & 14.96 & 0.437 & 0.580
          & 28.19 & 0.804 & 0.058
          & \textbf{29.71} & \textbf{0.833} & \textbf{0.057} \\
    \cmidrule{2-12}
      & \multirow{2}{*}{JPEG Compression}
          & \textit{QF}=90
              & 13.51 & 0.196 & 1.113
              & \textbf{22.64} & 0.554 & 0.318
              & 22.62 & \textbf{0.583} & \textbf{0.218} \\
      & & \textit{QF}=70
              & 13.46 & 0.190 & 1.261
              & \textbf{21.06} & 0.489 & 0.355
              & 19.81 & \textbf{0.522} & \textbf{0.292} \\
    \cmidrule{2-12}
      & \multirow{2}{*}{Gaussian Noise}
          & $\rho$=0.01
              & 18.97 & 0.582 & 0.393
              & 28.58 & 0.776 & 0.072
              & \textbf{31.62} & \textbf{0.864} & \textbf{0.048} \\
      & & $\rho$=0.04
              & 18.75 & 0.568 & 0.418
              & 22.58 & 0.551 & 0.208
              & \textbf{28.51} & \textbf{0.786} & \textbf{0.087} \\
    \cmidrule{2-12}
      & \multirow{2}{*}{Contrast Adj.}
          & $\eta$=0.9
              & 13.75 & 0.428 & 0.579
              & 31.68 & \textbf{0.914} & \textbf{0.017}
              & \textbf{32.73} & 0.908 & 0.043 \\
      & & $\eta$=0.8
              & 13.30 & 0.421 & 0.594
              & 26.30 & 0.835 & \textbf{0.045}
              & \textbf{30.72} & \textbf{0.845} & 0.059 \\
    \bottomrule[1.2pt]
  \end{tabular}
  \end{small}
  \caption{Stego image quality under different embedding capacities and five attack conditions. $\uparrow$ higher is better, $\downarrow$ lower is better.}
  \label{tab9}
\end{table*}

\begin{table*}[htbp]
  \centering
  \begin{small}          
  \begin{tabular}{@{}l l l l l l l l l l l l l@{}} 
    \toprule[0.8pt]
    \multirow{2}{*}{Capacity} & \multirow{2}{*}{Attack} & \multirow{2}{*}{Factor} &
      \multicolumn{3}{c}{Kishore et al.} &
      \multicolumn{3}{c}{Li et al.} &
      \multicolumn{3}{c}{Ours} \\[-1pt]   
    \cmidrule(lr){4-6}\cmidrule(lr){7-9}\cmidrule(lr){10-12}
    & & & PSNR$\uparrow$ & SSIM$\uparrow$ & LPIPS$\downarrow$
        & PSNR$\uparrow$ & SSIM$\uparrow$ & LPIPS$\downarrow$
        & PSNR$\uparrow$ & SSIM$\uparrow$ & LPIPS$\downarrow$ \\
    \midrule
\multirow{8}{*}{1.5 bpp}
  & Image Scaling      & $s{=}0.95$ & 12.26 & 0.209 & 0.673 & 28.14 & 0.859 & 0.074 & \textbf{33.86} & \textbf{0.948} & \textbf{0.018} \\
  \cmidrule{2-12}
  & Image Rotation     & $\sigma{=}0.1$ & 16.74 & 0.528 & 0.417 & 26.54 & 0.817 & 0.125 & \textbf{34.14} & \textbf{0.951} & \textbf{0.015} \\
  \cmidrule{2-12}
  & \multirow{2}{*}{JPEG Compression}
    & \textit{QF}=90 & 12.56 & 0.299 & 0.607 & 28.44 & 0.859 & \textbf{0.065} & \textbf{29.28} & \textbf{0.861} & 0.070 \\
  & & \textit{QF}=70 & 11.88 & 0.239 & 0.658 & 25.73 & 0.811 & \textbf{0.096} & \textbf{27.00} & \textbf{0.813} & 0.112 \\
  \cmidrule{2-12}
  & \multirow{2}{*}{Gaussian Noise}
    & $\rho{=}0.01$ & 23.13 & 0.749 & 0.140 & 31.29 & 0.905 & 0.040 & \textbf{32.04} & \textbf{0.920} & \textbf{0.037} \\
  & & $\rho{=}0.04$ & 15.90 & 0.517 & 0.392 & 26.27 & 0.811 & \textbf{0.101} & \textbf{27.44} & \textbf{0.816} & 0.124 \\
  \cmidrule{2-12}
  & \multirow{2}{*}{Contrast Adjustment}
    & $\eta{=}0.9$ & 15.05 & 0.440 & 0.478 & 17.41 & 0.562 & 0.382 & \textbf{34.62} & \textbf{0.968} & \textbf{0.016} \\
  & & $\eta{=}0.8$ & 13.32 & 0.323 & 0.562 & 14.57 & 0.405 & 0.564 & \textbf{34.38} & \textbf{0.953} & \textbf{0.017} \\
\midrule
\multirow{8}{*}{6 bpp}
  & Image Scaling      & $s{=}0.95$ & 12.36 & 0.209 & 0.673 & 14.86 & 0.452 & 0.536 & \textbf{31.04} & \textbf{0.888} & \textbf{0.059} \\
  \cmidrule{2-12}
  & Image Rotation     & $\sigma{=}0.1$ & 14.20 & 0.343 & 0.576 & 26.54 & 0.818 & 0.1245 & \textbf{29.81} & \textbf{0.885} & \textbf{0.054} \\
  \cmidrule{2-12}
  & \multirow{2}{*}{JPEG Compression}
    & \textit{QF}=90 & 11.65 & 0.227 & 0.690 & 19.53 & 0.686 & 0.263 & \textbf{23.60} & \textbf{0.720} & \textbf{0.253} \\
  & & \textit{QF}=70 & 11.45 & 0.218 & 0.699 & 17.54 & \textbf{0.617} & \textbf{0.362} & \textbf{19.24} & 0.572 & 0.411 \\
  \cmidrule{2-12}
  & \multirow{2}{*}{Gaussian Noise}
    & $\rho{=}0.01$ & 16.35 & 0.495 & 0.451 & 28.85 & 0.851 & \textbf{0.083} & \textbf{30.07} & \textbf{0.855} & 0.117 \\
  & & $\rho{=}0.04$ & 15.31 & 0.461 & 0.482 & 22.23 & 0.723 & 0.204 & \textbf{26.94} & \textbf{0.751} & \textbf{0.203} \\
  \cmidrule{2-12}
  & \multirow{2}{*}{Contrast Adjustment}
    & $\eta{=}0.9$ & 14.50 & 0.375 & 0.551 & 16.79 & 0.531 & 0.496 & \textbf{32.79} & \textbf{0.919} & \textbf{0.030} \\
  & & $\eta{=}0.8$ & 13.95 & 0.313 & 0.597 & 15.15 & 0.453 & 0.595 & \textbf{30.67} & \textbf{0.898} & \textbf{0.043} \\
    \bottomrule[1.2pt]
  \end{tabular}
  \end{small}
  \caption{Recovered secret image quality under different embedding capacities and five attack conditions. $\uparrow$ higher is better, $\downarrow$ lower is better.}
  \label{tab10}
\end{table*}

\bigskip

\end{document}